\documentclass[12pt,preprint]{aastex}

\shorttitle{Ghosts and Lumps in the Milky Way}
\shortauthors{Newberg et al.}

\begin{document}

\title{The Ghost of Sagittarius and Lumps in the Halo of the Milky
Way}

\author{
Heidi Jo Newberg\altaffilmark{1, \ref{RPI}},
Brian Yanny\altaffilmark{1, \ref{FNAL}},
Connie Rockosi\altaffilmark{\ref{UChicago}},
Eva K. Grebel\altaffilmark{\ref{MPH}},
Hans-Walter Rix\altaffilmark{\ref{MPH}},
Jon Brinkmann\altaffilmark{\ref{APO}},
Istvan Csabai\altaffilmark{\ref{EOT}},
Greg Hennessy\altaffilmark{\ref{USNO}},
Robert B. Hindsley\altaffilmark{\ref{USNO}},
Rodrigo Ibata\altaffilmark{\ref{STRA}},
Zeljko Ivezi\'{c}\altaffilmark{\ref{PU}},
Don Lamb\altaffilmark{\ref{UChicago}},
E. Thomas Nash\altaffilmark{\ref{FNAL}},
Michael Odenkirchen\altaffilmark{\ref{MPH}},
Heather A. Rave\altaffilmark{\ref{RPI}},
D. P. Schneider\altaffilmark{\ref{PSU}},
J. Allyn Smith\altaffilmark{\ref{UWO}},
Andrea Stolte\altaffilmark{\ref{MPH}},
Donald G. York\altaffilmark{\ref{UChicago}}
}

\altaffiltext{1}{Equal first authors}

\altaffiltext{2}{Dept. of Physics, Applied Physics and Astronomy, Rensselaer
Polytechnic Institute Troy, NY 12180; heidi@rpi.edu\label{RPI}}

\altaffiltext{3}{Fermi National Accelerator Laboratory, P.O. Box 500, Batavia,
IL 60510; yanny@fnal.gov\label{FNAL}}

\altaffiltext{4}{Dept. of Astronomy and Astrophysics, University of Chicago, 5640 S. Ellis Ave., Chicago, IL 60637\label{UChicago}}

\altaffiltext{5}{Max Planck Institute for Astronomy, KonigStuhl 17, D-69117 Heidelberg, Germany
\label{MPH}}

\altaffiltext{6}{Apache Point Observatory, P. O. Box 59, Sunspot, NM 88349-0059\label{APO}}

\altaffiltext{7}{Department of Physics and Complex Systems, Eotvos University, Pazmany Peter setany 1/A Budapest, H-1117, Hungary.\label{EOT}}

\altaffiltext{8}{Department of Astronomy, University of Arizona 933 N Cherry, Tucson, AZ 85726\label{UA}}


\altaffiltext{9}{Max-Planck-Institute fur Astrophysik Karl-Schwarzschild-Str. 1 D-85741 Garching, Germany \label{MPG}}

\altaffiltext{10}{US Naval Observatory, Washington, D. C.
\label{USNO}}

\altaffiltext{11}{University of Strasbourg, 67000 Strasbourg, France.\label{STRA}}

\altaffiltext{12}{Princeton University Observatory, Princeton, NJ 08544.
\label{PU}}

\altaffiltext{13}{Department of Astronomy and Astrophysics, The Pennsylvania State University, University Park, PA 16802\label{PSU}}

\altaffiltext{14}{Department of Physics and Astronomy, University of Wyoming,P. O. Box 3905 Laramie, WY 82071\label{UWO}}

\begin{abstract}
We identify new structures in the halo of the Milky Way Galaxy from positions, colors and magnitudes of five million
stars detected in the Sloan Digital Sky Survey.  Most of these stars are within $1.26^\circ$
of the celestial equator.  We present color-magnitude diagrams (CMDs) for stars in two previously
discovered, tidally disrupted structures.  The CMDs and turnoff colors are consistent with
those of the Sagittarius dwarf galaxy, as had been predicted.  In one direction, we are even able
to detect a clump of red stars, similar to that of the Sagittarius dwarf, from stars spread across
$110$ square degrees of sky.  Focusing on stars with the colors of F turnoff 
objects, we identify at least five additional overdensities of stars.

Four of these may be pieces of the same halo structure, which would cover a region of
the sky at least $40^\circ$ in diameter, at a distance of 11 kpc from the Sun (18 kpc from the center of the
Galaxy).  The turnoff is significantly bluer than that of thick disk stars, and closer to the
Galactic plane than a power-law spheroid.  We suggest two models to explain this new structure. 
One possibility is that this new structure could be a new dwarf satellite of the Milky Way, hidden in the Galactic
plane, and in the process of being tidally disrupted.  The other possibility is that it could be part
of a disk-like distribution of stars which is metal-poor, with a scale height of approximately 2 kpc
and a scale length of approximately 10 kpc.  

The fifth overdensity, which is 20 kpc away,
is some distance from the Sagittarius dwarf streamer orbit and is not 
associated with any known structure in the Galactic plane.  We have tentatively
identified a sixth overdensity in the halo.  If this sixth structure is instead part of a smooth distribution
of halo stars (the spheroid), then the spheroid must be very flattened, with axial ratio $q = 0.5$.  
It is likely that there are many smaller streams of stars in the Galactic halo.
\end{abstract}


\section{Introduction\label{intro}}


There is a growing body of evidence which shows that at least part of
the halo of the Milky Way Galaxy was formed through the accretion of
smaller satellite galaxies, and is not a relic of the initial collapse
of the Milky Way.  In the last decade, studies have convincingly
identified moving groups and substructure in the halo by identifying
groups of stars which are coherent in velocity \citep{mmh96,hwzz99}.
Simulations have predicted the existence of many halo streamers
\citep{ll95,jhb96,jzsh99,jsh99}.  Most recently, studies have
identified halo substructure and tidal stripping through spatial
information alone \citep{ietal00,ynetal00,oetal01}.  A striking example of
substructure in the halo is the identification of the Sagittarius
dwarf galaxy \citep{igi94}, and its associated stream of tidally
stripped stars, which appears to circle the Galaxy 
\citep{jsh95,iwgis97,il98,jmsrk99,hw01,magc01}.

The detection of substructure in the halo is important for our understanding
of the formation of our galaxy, and also as a test of cold dark matter
(CDM) and hierarchical clustering scenarios for structure formation in
the Universe.  For example, \citet{bkw01} argue
that the CDM scenario generically predicts large numbers of tidally disrupted
streams in the halo of the Milky Way - perhaps enough to account for the
stellar halo in its entirety.  They also suggest that the amount of halo
substructure could distinguish among proposed solutions to the ``dwarf
satellite problem,'' the tendency of CDM N-body simulations to predict
too many satellites in the halos of galaxies like the Milky Way and M31
\citep{klypin99,moore99}.

In \citet{ynetal00} (hereafter Paper I), we used a large sample of
faint blue stars from the Sloan Digital Sky Survey (SDSS) to discover 
two diffuse structures of stars in the halo.  Their inferred density 
indicated to us that these structures were disrupted remnants of a
previously bound structure, such as a dwarf galaxy.  We were not able
to see the full extent of either structure.  \citet{iils01}
explained these two structures as two slices through the same great
stream which completely circles the galaxy. 
The positions and distances of the stars in our structures exactly
matched those expected from the tidal disruption of the Sagittarius dwarf
spheroidal galaxy.
Other pieces of this same stream have been recently reported by \citet{detal01}
and a simple model of the Sagittarius breakup is given in \citet{hw01}.

In this paper, we present additional observations of the equatorial
($-1.26^\circ < \delta_{2000} < 1.26^\circ$) data from the SDSS
which probe a significantly larger angle of right ascension along the equatorial
ring than in Paper I.  We extend the methods of Paper I, which used 
faint blue stars with A-type colors to trace structure, to include the 
much larger sample of turnoff or near turnoff stars with F-type colors.  
These new data contain strong evidence for further halo substructure.
The key figure in this paper is a 2D polar density histogram $(\theta,r) = (\rm RA, g')$ of stars in the plane of the celestial equator with
F colors (primarily F dwarf stars)
shown in Figure~\ref{Fwedge} and described in detail in \S5.  


We expect to detect these streams of stars in addition to, or as part of, the
individual stellar components of the Milky Way galaxy.  \citet{bs84}
published the ``standard galaxy model," which contained two components: a
thin disk modeled with a double exponential profile with scale height 0.325 kpc,
and a halo modeled with a slightly flattened power-law spheroid with axial
ratio 0.80.  In the solar neighborhood, the spheroid stars were outnumbered
by the thin disk stars by a factor of 1 in 500.  An additional component,
a thick disk, was proposed by \citet{gr83}.  Since then, the popularity
of models with a thick disk component has grown.  The thick disk is typically
modeled as a double exponential with a scale height of about $1 \pm 0.5$ kpc and a
stellar frequency, compared with the thin disk, in the solar neighborhood of 
between 1:8 and 1:50 \citep{rr01,rm93,obrcm96,rhcob96,brk99,cetal01,kjs01}.  
Stars in the thick disk component dominate the star
counts 2 to 5 kpc above the plane, and have chemical and kinematic properties
intermediate to the thin disk and halo populations.  See \citet{n99} for a
review of the status of the thick disk.  See  \citet{gwk89};
\citet{m93}; and \citet{w99} for reviews of Galactic components.  

The literature on the subject
of Galactic components is vast; studies include star count analyses, kinematics,
chemical properties of stars, and comparisons with other galaxies.  We have summarized
only the most basic structures, which may themselves have substructure, and which 
some authors may break into parts or name
differently.  We have not discussed stellar populations in the Galactic center, such
as the bulge population.  See \citet{f88}, \citet{f99} for reviews of the Galactic bulge.

\section {Observations} 

The observations are from several time-delay and integrate (TDI) CCD scans 
obtained under photometric conditions in good seeing (FWHM $< 1.9''$) on
twelve nights between 1998 September 19 (run 94) and 2001 February 20 (run
2126) with the Sloan Digital Sky Survey (SDSS) mosaic imaging 
camera \citep{getal98}.  See \citet{yetal00} for a technical overview 
of the survey.

A single `run' scans six 0.23 degree wide swathes of sky (`scanlines') separated by
gaps of about 0.2 degrees.  The gaps are filled by a second `strip', containing
six scanlines, which completes a filled `stripe' on the sky.  The SDSS survey
area is divided into 48 numbered stripes, each 2.5$^\circ$ wide.  
Each stripe is an arc of 6 to 10 hours in length that follows a great
circle which passes through the survey poles, ($\alpha, \delta$) =
($275^\circ, 0^\circ$) and ($\alpha, \delta$) = ($95^\circ, 0^\circ$).
Equinox J2000 is implied throughout this paper.  Most of the survey area is in the North
Galactic cap.  The few stripes in the South Galactic cap have separate
stripe designations from their northern counterparts on the same great
circle.  In particular, the celestial equator ($\delta = 0$) is designated stripe 10 
above the Galactic equator, and stripe 82 below.
See \citet{setal01} for further details of the survey conventions.  

The star count work of this paper requires that the sampling of stars be
quite uniform over a large area of sky.  Because of the way the SDSS map of the
sky is obtained, and pieced together in a mosaic fashion,
it is important for the purposes of this paper to select
a sample which does not unintentionally `double count' objects in the boundary regions of
overlapping survey pieces.   
The full SDSS database contains multiple copies of many objects.  We select single copies
of the objects using several flags stored with the object in the database.  During image processing, objects are
extracted from each scanline one `field' at a time, where the breaks between fields
are imposed somewhat arbitrarily every 1361 rows.  So that objects which lie on these
breakpoints are not lost, overlaps are processed with adjacent `fields.'  Objects
which fall in an overlap may be in the catalog twice.  Also, there are overlaps between
the interleaved strips which make up a stripe.  The flag `OK\_SCANLINE' is assigned
to only one copy of each object in an individual scanline, and uses astrometric declination
limits (on the equator) to flag only the non-overlapping areas of the two strips in
each stripe.  If you take all objects from two interleaved runs which have
`OK\_SCANLINE' set, you will get one instance of each object from the combined two runs.  Two 
instances of the same object may both have OK\_SCANLINE set if they are in overlapping
runs covering exactly the same part of the sky; however, since the dataset used in this paper
was constructed with only non-overlapping portions of runs, selecting with the OK\_SCANLINE
flag produces only one instance of each object in a given stripe.  Objects can also lie 
on overlaps between different stripes.  

The flag `PRIMARY' is assigned to one copy
of each object in the entire database.  Each numbered stripe is assigned a region in the
sky over which its objects will be PRIMARY.  Each numbered `run' is assigned a region of
the stripe over which it is PRIMARY.  Since the stripes overlap more toward the survey 
poles, the area of sky over which the stripe is PRIMARY decreases towards its ends.  
To keep a sample of objects on a single stripe uniformly sampled
in declination, one selects those with the OK\_SCANLINE flag set (rather than the PRIMARY flag).

Most of the data used in this paper are located on stripes 10 and 82 on the
celestial equator ($-1.26^\circ < \delta < 1.26^\circ$).  We also use data from
stripe 11, 2.5$^\circ$ above the equator, stripe 12, at $\delta \sim
+5^\circ$, and stripe 37, which follows part of an arc of a great circle 
tilted 67.5$^\circ$
relative to the equator.  Table 1 presents details of the strips,
the stripes, and the sky coverage of the data used in this paper.  Not all
sections of the equator scanned have both strips filled.  In
particular the data from the ends of runs 752, 756 and 1755 don't have a
corresponding filling strip.  In order to have uniform star count
statistics at all azimuths in these cases, double copies of the single
strips were made to normalize the number counts to those areas of sky
where two filled stripes were available. This is indicated by a ``2''
in the multiplicity column of Table 1.  

The photometric system for the SDSS includes five filters, $u'\> g'\> r'\>
i'\> z'$ \citep{figdss96}.  The system is approximately $AB_{\nu}$ normalized, with central
wavelengths for the filters of 3543\AA , 4770\AA , 6231\AA , 7625\AA , and
9134\AA , respectively, and effective widths of typically 1000\AA .
Since the precise calibration for the SDSS filter system is
still in progress, magnitudes in this paper are quoted in the $u^* g^*
r^* i^* z^*$ system, which approximates the final SDSS system \citep{setal}.
These systems differ absolutely (with
small color terms) by only a few percent in $g^* r^* i^* z^*$,
and no more than 10\% in $u^*$.  

The data were reduced with PHOTO \citep{l01} versions 5.1 and 5.2, and 
astrometrically calibrated with the ASTROM pipeline described in \citet{pmk01}.

\section {Data Reduction\label{datared}}

The SDSS software generates a database of measured object parameters and 
flags, including information on deblended `children' of 
sources whose profiles overlap.
One must select from this database a list of interesting objects to be
used as input to analysis routines.  We selected from the photometric
catalog only those objects which were marked as stellar, unsaturated,
and not too near the edge of the frame (too near is generally about $8''$).  
In order to ensure that only one instance
of each object appears in the final object tables, we selected objects
which were marked as OK\_SCANLINE as explained above. 


Using these criteria, we generated a catalog of 4.3 million
stars to $g^* \sim 23.5$ on the equator. The total area covered
is approximately 560 square degrees.  Data on off-equatorial stripes
add an additional 0.7 million stars over approximately $70$ square degrees.
Completeness vs. magnitude is discussed below, but we note here that
for objects with $g^* > 22.5$, the star-galaxy separation results in
most stellar objects being classified as galaxies and these thus are not 
prominent in our subsample.

The SDSS software measures object flux in a variety of ways.  Since we
are measuring stars only, we use magnitudes calculated from a fit of
modeled stellar profiles (point-spread-function, or PSF, magnitudes) to 
each object.  We correct
these magnitudes for reddening using $E(B-V)$ from \citet{sfd98}, which
has spatial resolution of 0.1 degrees, and the
standard extinction curve \citep{ccm89}, which for SDSS filters yields: $A_{u^*} =
5.2 E(B-V); A_{g^*} = 3.8 E(B-V); A_{r^*} = 2.8 E(B-V)$.

The flux of objects are presented in an inverse hyperbolic sine
(asinh) representation of \citet{lgs99}.  This definition
has the feature (unlike a magnitude) that it is well defined for zero or negative fluxes, which
can result from measurement of no flux at the position of an object
detected in a different filter.  Asinh numbers and magnitudes are the
same to better than 0.1\% for objects with $g^* < 21$, differ
by 0.1 mag at $g^* = 23.9$.  At zero flux, the asinh numbers go through $g^* \sim 25$.  The $r^*$ asinh flux shifts in the same way as the $g^*$, and thus there is
negligibly little change in $g^*-r^*$ color due to using these asinh numbers
instead of magnitudes.  
For the magnitude ranges of 
interest here ($g^* < 22.5$), the difference
is unimportant.  In the remainder of this paper, we will refer to
asinh numbers as magnitudes.

Figure~\ref{redvsalpha} shows reddening ($10\times E(B-V)$ in magnitudes) 
from \citet{sfd98} around the celestial equator.  
We also plot number counts in 10 degree bins for a sample of color 
selected stellar objects with $18 < g^* < 21.5, 0 < u^*-g^* <0.3, 
0.1 < g^*-r^* < 0.3$ versus right ascension around the equator.  These 
objects, which are primarily quasars (see Figures 1, 4, and 5 of Paper I), 
should have a constant number density independent of Galactic 
latitude.  Figure~\ref{redvsalpha} shows that the selection of stars of similar colors and 
magnitudes as a function of $\alpha$ around the sky is mostly unbiased. 
Near $\alpha = 60^\circ$  the amount of intervening interstellar dust is quite
large and the errors in the reddening corrected magnitudes are
larger than at most other $\alpha$.
For $310^\circ < \alpha < 350^\circ$, only one of two SDSS strips of
data is present, and thus the counts have less S/N than the rest of
equatorial data.  The counts have been normalized upwards  by a factor of two
in the figure and the error bars appropriately increased.  Even with this normalization,
it is apparent that the counts fall systematically slightly below 
that of those at, for example, $150^\circ < \alpha < 230^\circ$.  We are 
uncertain of the reason for this.

From inter-comparison of objects detected twice in overlapping scans, we find the
rms error for stellar sources with $g^* < 19$ is typically $\sim 2$\%.  For
objects with $20 < g^* < 21$, typical errors are 5\%, growing to
20\% at $g^* = 23.5$ near the detection limit.  For reference, blue
stars with $0 < B-V < 0.2$ have an SDSS $g^*$ magnitude approximately
equal to their Johnson $V$ magnitude.  A theoretical color transformation 
is given by \cite{figdss96}: $g^*-r^* = 1.05(B-V) - 0.23$.

We plot in Figure~\ref{limmag} the rms dispersion of the difference in $g^*-r^*$ color
for matched objects between two runs as a function of magnitude.
For bright magnitudes this dispersion reflects the photometric errors of 
2\% in g and r (about $2\times \sqrt{2}\sim $3\% in uncorrelated color).
The photometric error increases to nearly 20\% for objects near $g^* \sim 22.5$.

For some of our analyses, it is crucial to know the limiting
magnitude, or more precisely the magnitude limit at which the survey
can be considered complete, as a function of color and position in the sky.
In the same Figure~\ref{limmag}, we show the fraction of stars matched between two
overlapping runs which make up the equatorial stripes.  The matched
fraction is calculated at two positions around the equator.  One
segment of matched data are at lower latitude, averaging $b \sim 30^\circ$
with $125^\circ < \alpha < 145^\circ$, while the other are at higher
latitude with average $b \sim 50^\circ$, and $145^\circ < \alpha <
230^\circ$.  Stars in three color ranges, $0.1 < g^*-r^* < 0.3, 0.3 <
g^*-r^* < 0.4$ and $0.6 < g^*-r^* < 0.7$ (all with $u^*-g^* > 0.4$) in
one strip are matched to the full list of stars in the overlapping
strip.  The fraction which are matched is recorded as a function of
$g^*$ magnitude.  Figure~\ref{limmag} shows that for all three color
bins of the high latitude matched set (the low latitude set is identical within
the errors), the matched fraction is
constant to about $g^* \sim 22.5$, after which it drops off steeply,
and somewhat more quickly for objects of bluer color.

One notes that for bright objects, the matching fraction is
not 100\%.  This is in part due to the fact that edge overlaps were used
between the interleaving stripes (so that the same exact area of sky is not
sampled by each overlapping patch).  Also, the reduction software doesn't
resolve all objects around bright stars into separate detections.
Independent matches against external catalogs indicate that
the detection software detects over 99\% of objects at these magnitudes ($18 < g^* < 22$).

Both segments of data at higher and lower latitudes give
the same results for $g^* < 22.5$.  Thus any selection we do based on 
magnitude $g^* < 22.5$ is free of significant color or completeness
bias to this limit.  The variation in density of objects with quasar colors 
indicates there is possibly some small variation in completeness limits or
problems with reddening corrections at 
$60^\circ < \alpha < 76^\circ$ and $318^\circ < \alpha < 325^\circ$. 

The imaging pipeline separates detected objects into stars and galaxies
based on goodness of fit to PSFs and model galaxy
profiles.  For the seeing conditions under which these data were obtained, 
this separation produces excellent results to approximately $g^* \sim 21$.  
We show in Figure~\ref{galaxies} a color magnitude image of $\approx
100,000$ objects typed as galaxies, selected around the celestial
equator, and binned as a Hess diagram \citep{h24}.  Nearly all galaxies
have colors redder than $g^*-r^* > 0.4$, significantly redder than the
turnoff stars we are interested in at $g^*-r^* \sim 0.3$.  There is
some leakage of stars into the galaxy population for $g^*-r^* \sim 0.3$
at $g^* > 22.5$, again below the limits set by Figure~\ref{limmag}.
The galaxy population's localization in color-magnitude space 
affects none of the conclusions made here about turnoff-color 
star counts.

\section {The Ghost of Sagittarius\label{ghostsag}}

Since the positions and colors of the giant branches and horizontal
branches of dwarf galaxy companions to the Milky Way differ considerably as
a function of the dwarf's metallicity, age and stellar population mix,
these features can be used as identifying signatures of a given dwarf
galaxy or cluster.  In this section, we explore the color-magnitude distribution
of stars in previously detected clumps.  This discussion will motivate our
use of F-colored stars to detect spatial structure in the next section.

Using a technique similar to that of \citet{mskrjtlp99}, we will construct
a color-magnitude diagram of the two concentrations 
of stars from Paper I.  
To avoid confusion in referencing overdensities of stars, we will name
them S$l\pm b-g$ where ($l\pm b$) are the Galactic
coordinates of the approximate center of the structure (where it
intersects an SDSS stripe), and $g$ is the approximate $g^*$ magnitude
of the `turnoff' F dwarf stars in that structure.  If the structure is
identified with a known halo component, such as the Sagittarius dwarf,
then the identification may appear in parentheses after the structure
name.  Under this naming convention, the two structures identified in
Paper I are given the names S341$+$57$-$22.5 and S167$-$54$-$21.5, and may
be clearly seen in Figure\ref{Fwedge}.

For structure S341$+$57$-$22.5, we used 
stars marked as `PRIMARY' in stripes 10
and 11 with $200^\circ < \alpha < 225^\circ$ and $u^*-g^* > 0.5$.  The cut
in $u^*-g^*$ eliminates blue quasars which would otherwise dominate
the faint blue edge of the color-magnitude diagram.  
The `PRIMARY' stars
come from non-intersecting portions of stripes.  Since the stripes are parts of great 
circle arcs, the non-overlapping portion of the
stripe is thinner towards the survey poles than it is on the survey
equator.  On the celestial equator at $\alpha = 200^\circ$, the width of
stripe 10 and 11 together is 4.8 degrees.  On the celestial equator at
$\alpha = 225^\circ$, the width of the two stripes is 3.8 degrees.  The total
area covered is 110 square degrees.  

Due to the large number of stars in this area of sky, we generated an
image of counts-in-cells of the color-magnitude diagram, with a bin
width of 0.02 in $g^*-r^*$ and 0.05 in $g^*$.  In order to reduce the
number of field stars in the image, we subtracted a
similarly generated color-magnitude image of stars in a similar portion of the
sky which does not contain the Sagittarius dwarf.  The subtracted stars
are from stripe 10 and 11, $170^\circ < \alpha < 180^\circ$, plus twice
stripe 10, $230^\circ < \alpha < 235^\circ$ (SDSS has not yet processed
data in the $230^\circ < \alpha < 235^\circ$ range for stripe 11).

For stars at S341$+$57$-$22.5, the resulting Hess color-magnitude diagram image (with a greyscale stretch 
proportional to the square root of the number of stars in each bin) 
is shown in Figure~\ref{sagsubn}.  One can clearly see the turnoff at ($g^*-r^*,g^*)= (0.2,22.5)$, 
the giant branch at $(g^*-r^*,g^*) = (0.5, 22)$ running to $(g^*-r^*,g^*) = (0.6, 20.5)$, blue stragglers
at ($g^*-r^*, g^*)=(-0.1, 21.5)$, a 
blue horizontal branch at $(g^*-r^*,g^*)=(-0.15, 19.2)$ and a clump of red stars
at ($g^*-r^*, g^*)=(0.55, 19.6)$.  For comparison,
we show in Figure~\ref{sagcmd} the identical plot for the
Sagittarius dwarf itself, with data from \citet{mbcimpp98}.  Since these
Sagittarius dwarf data were taken with V and I filters, and the dwarf is much
closer than the dispersed clump (23 vs. 45 kpc), we arbitrarily
aligned the clump of red stars and applied a linear correction in the color direction
so that the distance between the clump of red stars and the point where the
horizontal branch and the upper main sequence (as traced by blue stragglers) meet is the same in each
image.  The adopted transformation equations are: $g^* = V + 1.39$ and
$(g^*-r^*) = 0.9(V-I) - 0.41$.  The relation between $g^*$ and $V$
includes both the difference in distance modulus and the filter
transformation.  Any differences in reddening or errors in reddening correction for either
data set are implicitly included in the transformation.  We then compared this
empirical transformation with one derived from theoretical SDSS filter curves
as given in \citet{figdss96}, and find that they match to within about 5\%.
The distance compensation, $g^* = V + 1.39$ is also within a few percent
of the expected theoretical value, $g^* = V + 5 log(45/23) = V + 1.45$.

The similarity of color-magnitude diagrams can be judged from the color
of the turnoff, the distance from the horizontal branch to the turnoff,
the slope and degree of population of the red giant branch, the presence or
absence of blue stragglers, and the color distribution of stars along the horizontal
branch.  Though there are some differences (most notably - the Sagittarius dwarf photometry
shows few if any blue horizontal branch stars), the agreement between the color magnitude 
diagrams for the Sagittarius dwarf and our 110 square degree patch of sky on
the equator is striking, and leaves little doubt that the tidally
disrupted clumps of stars discovered in \citet{ietal00} and Paper I
are in fact pieces of the Sagittarius dwarf stream, in exactly the
positions predicted by \citet{ilitq01}.

It is interesting to estimate the fraction of the Sagittarius dwarf
which is present in our observed piece of its orbit.  One measure is
the number counts in the clump of red stars.  We estimate that there are $500
\pm 50$  stars in the 110 square degree
patch of sky with $19.30 < g^* < 19.65$ and $0.52 < g^*-r^* <
0.66$.  For this estimate, we measured the clump of red stars in the 
unsubtracted color-magnitude image to reduce the statistical noise in 
the measurement (the background was determined by linear interpolation).  \citet{igi95}
find $17,000$ horizontal branch stars in a
patch of sky thought to contain half the mass of the Sagittarius
dwarf.  We detected about $500/34,000 = 1.5\%$ as many red stars in the clump
as are present in the dwarf itself in a portion of its orbit extending
$4.4^\circ$ on the sky.  The orbit is roughly perpendicular to our scan
line and presumably extends over $360^\circ$ on the sky.  If the stellar
density along the stream were constant (admittedly a naive assumption), and
the stream only wraps around on itself (so as not to produce multiple streams at other positions on the sky), then this implies about as many stars (1.2 times as many in
this calculation) in the stream as in the undisrupted dwarf. 
This number is interesting, but has a large error bar as the fraction of
stars in the clump of red stars could easily differ between or within
the dwarf and the stream.

Using exactly the same procedure as for S341$+$57$-$22.5 (Sagittarius), we
generate a color-magnitude image for S167$-$54$-$21.5, which has
also been tentatively identified as a piece of the Sagittarius stream
by \citet{iils01}.  Since we have no data adjacent to stripe 82, we
used the full width of the collected data (there is no change in the
$\delta$-width of the stripe on the sky as a function of right ascension).  We
used all stars with $15^\circ < \alpha < 50^\circ$ to make the clump
color-magnitude image.  We then subtracted a color-magnitude image of
all stars with $10^\circ < \alpha < 15^\circ$ plus double-counting all of
the stars with $0^\circ < \alpha < 5^\circ$ and $50^\circ < \alpha <
55^\circ$.  The resulting color-magnitude image is shown in
Figure~\ref{sagsubs}.  One can see a clear turnoff, a giant branch,
and blue straggler stars.  The horizontal branch and clump of red stars, though
possibly faintly present, are not compelling.  There are definitely
blue horizontal branch stars present, since this
clump was originally detected in A-colored stars in Paper I.  The
number of blue horizontal branch stars in the Sagittarius south stream is 
only a third the number in the northern stream.  This color-magnitude 
diagram is consistent with that of the Sagittarius dwarf, 
but does not present as
strong a case for identification as that in the North.

\section {Halo structure in F stars}

\subsection {Distribution on the celestial equator}

Our detection of halo structure in Paper I 
relied on the standard candle
characteristics of A-colored blue horizontal branch stars.
The results of the previous section demonstrate our ability to examine
the structure of the Milky Way using stars as faint as F dwarfs.  
If it is possible to use the much larger numbers
of these main sequence stars, one 
anticipates that much more tenuous halo structures
could be discerned, though not as far out
into the halo.  We pursue such a path.

From stripes 10 and 82, we generate a catalog of $4,270,645$
stars with $u^*-g^* > 0.4$ and $-1.0 < g^*-r^* < 2.5$.  The $g^*-r^*$
color range is wide enough to include essentially all stars.  
In this section, we
will be using not just the PRIMARY flagged stars, but all of the OK\_SCANLINE
flagged stars from the runs used to fill in stripes 10 and 82 as detailed in Table 1.  
This way, the width of
the stripe in declination does not change as a function of right
ascension.  The $u^*-g^*$ color cut removes primarily low redshift QSOs.

In Figure~\ref{s10230240cmd} we show a color-magnitude image of all of
the stars centered in the direction (l,b) = (5,40), $230^\circ <
\alpha < 240^\circ$ in stripe 10.  The stars with $g^*-r^* \sim 0.5$ and 
$g^* \sim 18$ are thought to be associated with the thick disk of the Milky Way.  The
stars with $g^*-r^* \sim 1.3$ are M stars in the thin disk, the thick
disk, and, at $g^* > 22$, the halo.  The bluer stars ($g^*-r^* \sim 0.3$)
are generally ascribed to the halo.  The clear separation in turnoff color between
the ``thick disk" and ``halo" was described by \citet{cetal01}.  The stars
we are interested in are the bluer stars with $g^*-r^* \sim 0.3$, at
$g^* > 19$, which are associated with the halo population.
It is important to note that using a color separation to distinguish
between ``thick disk'' and ``halo'' populations, though it appears
to work well, is an empirical one, and is a separate distinction from 
a kinematic separation of the populations. 

To separate the thick disk stars from halo stars, we select
$334,066$ stars in stripes 10 and 82 (which are both on the celestial
equator) with $0.1 < g^*-r^* < 0.3$, keeping the $u^*-g^* > 0.4$ color cut.  This
cut includes only the bluer ``halo" stars; we have so
many stars that we have the luxury of throwing half of them away to
reduce thick disk contamination and keep a much smaller range of dwarf
star absolute magnitudes in the sample.  We plot this sample of stars
in a 2D polar density histogram in Figure~\ref{Fwedge}.  This figure is similar to
the wedge plots of Paper I (see Figure 3 of that paper), but is
displayed in an image by binning all the stars that would have
appeared as individual dots within each pixel.  The Sun
is located at the center of the plot.  Stars of the same apparent magnitude
are at the same radial distance from the center of the plot, with
$g^*=11$ at the center of the plot and $g^*=24$ at the edge (though the
data cuts off at $g^* = 23.5$).  If each
star has the same intrinsic magnitude (roughly the magnitude of an F main sequence
star), then the radius from the center of the diagram scales as the
logarithm of the distance from us.
Typical distances probed with turnoff stars range from a few kpc to
about 60 kpc at the edge of the plot ($g^* = 23$).

The shading of each box indicates the relative number of F stars within
the pixel's azimuth and magnitude ranges, with 7.69 pixels/magnitude.  It is
generated by calculating for each star in the sample the (x, y) position
the star should go in Figure~\ref{Fwedge}, and then incrementing the count in the pixel which covers that spot.

Figure~\ref{Fwedge} does not show the smooth distribution of stars
expected from a power-law spheroid or exponential disk stellar density 
distribution.  The overdensities of stars at $(\alpha, g^*) = (210^\circ, 22)$
and $(\alpha, g^*) = (40^\circ, 21)$ are  F dwarfs associated with
S341$+$57$-$22.5 and S167$-$54$-$21.5 (Sagittarius).  The dark radial line at
$\alpha = 229^\circ$ is the main sequence turnoff of the globular
cluster Pal 5.  

The feature at $\alpha = 60^\circ$ is exactly coincident with a large interstellar
dust cloud at that position in the sky, and does not represent halo
structure.  When the reddening correction is that large, one must worry about the
distance to the source(s) of reddening, and the accuracy of the maps.  Small
differences in applied reddening change the intrinsic colors of the selected
objects.  The counts in this direction are consistent with over-correction for
reddening, which moves redder stars into the color selection box.
As is apparent in the color-magnitude Hess diagrams, the redder turnoff stars
are more prevalent at brighter magnitudes.

Locations of other interesting overdensities are
labeled in Figure~\ref{Fwedge}
and summarized in Table 2. These and other overdensities will be discussed in detail below.

\subsection {Spheroid models\label{spheroidmodels}}

What do typical Galactic stellar component models, such as an exponential thick disk or
a power-law spheroid, look like in a wedge image such as that of Figure~\ref{Fwedge}?
 We will use the term `spheroid'
to describe any smooth distribution of stars (in excess of the known
thin and thick disk populations) in the halo of the Milky
Way, regardless of its density profile.  The halo of the Milky Way is
a region of space containing gravitationally bound matter.
The halo stars of the Milky Way are a combination of dwarf 
galaxies, globular clusters, streamers, and a smooth component (spheroid).  
The usual density profile for the
Galactic spheroid is a power-law, or alternatively a flattened power
law with flattening parameter $q$, given by:
\[\rho = \rho_o (X^2 + Y^2 + Z^2/q^2)^{\alpha/2},\]
where X, Y, and Z are the usual Galactocentric coordinates with Z
perpendicular to the Galactic plane, and $\rho_o$ sets the density scale.
If $q=1$, the model is spherically
symmetric.  $\alpha$ is thought to be about $-3.2\pm 0.3$ (see Paper I).

To generate the wedge image, we must transform to a heliocentric coordinate
system, $(l, b, R)$, where $R$ is the distance from the Sun.  In this coordinate
system, the number of stars per magnitude bin is given by:
\[\frac{dN}{dm} = \frac{dR}{dm} \frac{dN}{dR} = (\frac{R}{5}) (\Omega R^2 \rho_o r^\alpha),\]
where 
\[r^2 = R_o^2 + Q R^2 - 2 R_o R \cos(l) \cos(b),\]
\[Q= \cos^2(b)  + \sin^2(b) /q^2.\]

In our simulations, we assume the distance to the center of the
galaxy, $R_o$, is 8.0 kpc.  In our plots, the number of pixels in a
given apparent magnitude annulus of width $dm$ is proportional to $(m-11)$,
and the width of the data in declination is constant.  Therefore,
\[\Omega \propto (m-11)^{-1},\]
as $dm$ is approximately constant for each pixel in the wedge image of Figure~\ref{Fwedge}.  In this
way we correct for the angular size of each pixel.

In order to construct the simulation, we need to relate the distance $R$ to
the apparent magnitude $m$.  For this we need to know the approximate
absolute magnitude of the stars in Figure~\ref{Fwedge}.  Clearly,
there will be a spread in stellar magnitudes, which should result
in a broader distribution in the data than in the simulated image.  We
find an estimate of the magnitudes of turnoff stars by using the
distance in magnitudes from the horizontal branch of the S167$-$54$-$21.5
(Sagittarius, from Paper I) to the turnoff of the Sagittarius stars in the
Figure~\ref{Fwedge}, stripe 82.  Figure~\ref{Fstarmag} shows the distribution of
apparent magnitudes for stars with $30^\circ < \alpha < 45^\circ$.
In each magnitude bin, we have subtracted the number of stars in
a similar region of the equator that does not include the Sagittarius
stream ($20^\circ < \alpha < 25^\circ$ and $45^\circ < \alpha < 55^\circ$).
The range of apparent magnitudes at the peak of the distribution is
$21.1 < g^* < 21.8$.  Assuming a horizontal branch absolute magnitude $M_{g^*} \sim 0.7$
and $g^* = 18$ (Paper I), the absolute magnitudes of the stars in the image are
estimated to be in the $3.8 < M_{g^*} < 4.5$ range, quite typical for
F dwarfs.  We adopt $M_{g^*} = 4.2$ as the typical magnitude of a turnoff star.
This value is consistent with that estimated from SDSS magnitudes and known
distances of the globular cluster Palomar 5. 

We would like to draw your attention to some special azimuthal
directions on the wedge plot of Figure~\ref{Fwedge}.
The Galactic plane intersects the plane of the plot at
$\alpha = 103^\circ$, and goes straight through the
plot center to $\alpha = 283^\circ$.  The place where $l=0^\circ$ is at
$\alpha = 228^\circ$, almost in the direction of Pal 5 at $\alpha =
229^\circ$.  In the celestial equatorial plane, this is not the
direction of the Galactic center, but above the Galactic center in the
direction which will intersect the Z-axis of the Galaxy.  If the
Galactic spheroid were very prolate ($q \rightarrow \infty$),
then the highest density of stars intersected by the celestial equator
would be at $l = 0^\circ$.  If the spheroid were flattened into a pancake
($q \rightarrow 0$), the highest density of stars would be in the
Galactic plane at $\alpha = 283^\circ$ ($b=0^\circ, l=32^\circ$), and there would be
another high density of stars at $\alpha = 103^\circ$ ($b=0^\circ, l=212^\circ$);
the relative densities would depend on how quickly the power-law drops
off.  If the spheroid is spherical, the highest density will be in the
direction of the closest approach to the center of the Galaxy.  This
direction does not depend on the slope of the power-law or the
inferred absolute magnitude of the stars.  We show in Figures~\ref{models}a,
\ref{models}b, and \ref{models}c images of the wedge image simulations which result from $q = 0.5, 1.0,$ and
$1.5$ power-law spheroid models with $\alpha = -3.5$.

Using a similar procedure to that for the spheroidal models, we can
generate a simulated exponential disk as seen in the cross
section of the celestial equator.  The relevant equations for the disk
density profile models are:
\[\rho = \rho_o e^{-r/s_l} e^{- |Z|/s_h},\]
\[\frac{dN}{dm} = \frac{dR}{dm} \frac{dN}{dR} = (\frac{R}{5}) (\Omega R^2 \rho_o e^{-r/s_l} e^{-|Z|/s_h}),\]
where $r^2 \equiv X^2 + Y^2$ and X, Y, Z are standard Galactocentric
coordinates with the Sun at (X,Y,Z) = (-8.0,0,0).  $\Omega$ is the
same as in the spheroid model.  A simulated exponential disk with scale 
length $s_l = 3$ kpc and scale height $s_h = 1$ kpc is shown in Figure~\ref{models}d.  We used the same
absolute magnitude for the simulated stars.  Exponential disks generally put
concentrations of stars at $b = 0$ ($s_h << s_l$).  They result in
concentrations of stars at $l = 0$ if $s_h >> s_l$.

\subsection{Overdensity at $\alpha = 190^\circ$ - S297$+$63$-$20.0}

Armed with these results, we turn our attention again to the data in
the wedge plot of Figure~\ref{Fwedge}.  Neither an exponential disk
model nor a power-law model can put a density peak at
$180^\circ < \alpha < 195^\circ$, where $(l,b)=(297^\circ,63^\circ)$.  We
tentatively identify the concentration at about $g^* = 20.5$ in
Figure~\ref{Fwedge} as a stream or other diffuse concentration of
stars in the halo, and name it S297$+$63$-$20.0.  
Figure~\ref{image-S10-180-195} shows the color-magnitude diagram for stars
with $180^\circ < \alpha < 195^\circ$.

A recent paper by \citet{v01} present corroborating
evidence for this stream from observations of 5 clumped
RR Lyraes at $\alpha=197^\circ$, at a similar inferred distance from
the Sun (20 kpc).

\subsection {Overdensity near $\alpha = 125^\circ$ - S223$+$20$-$19.4}

We now turn our attention to the concentration of stars near $\alpha
= 125^\circ$ at $g^* \sim 19.5$. As we noted above, it is possible to put concentrations
of stars near the Galactic plane near the anti-center
($l \sim 180^\circ$) with either an exponential disk or a flattened 
spheroidal power-law model ($q < 0.6$).  These stars are too faint to be produced
by thin disk or thick disk stars from the double exponential profiles of any
standard models.  As we will show in \S5.7, it would
be necessary to postulate an unusually flattened power-law distribution,
or an unusually large scale length exponential disk distribution, to
put enough stars this far away from the Galactic center
and still fit the number counts towards the Galactic center.

A color-magnitude image
for stars in this direction is shown in Figure~\ref{cmd125}.  
Most of the brighter, bluer stars ($g^* < 18.5$,
$0.4 < g^*-r^* < 0.6$), presumably thick and thin disk, are part of a
distribution with a redder turnoff than the fainter stars at ($g^*
> 20$).  What is stunning about the color-magnitude image in this
direction is that the fainter, bluer stars appear to follow a main
sequence, as if the stars are all at about the same distance from the
Sun.  In any kind of exponential disk or power-law
distribution, one expects a much broader distribution of distances,
which spreads the stars in the vertical direction on the
color-magnitude diagram.
See Figure 6a of \citet{ls00} for an example of
how a dwarf spheroidal in the field looks in such a CMD.

We show the shallow depth of the structure quantitatively in Figure~\ref{dwarfthickness}.
Figure~\ref{dwarfthickness}(a) shows power-law models for a variety of
slopes and flattenings in the direction $(\alpha,\delta) = (125^\circ,0^\circ)$.  
Figure~\ref{dwarfthickness}(b) shows a
variety of exponential disks in the same direction.  Compare the widths of the peaks of
these models with the width (in magnitude) of the main sequence at
$\alpha \sim 125^\circ$ shown in Figure~\ref{dwarfthickness}(c).  The
black line gives star counts vs. magnitude in the color range $0.1 <
g^*-r^* < 0.3$. Note the peak centered at $g^* = 19.4$.  
If the data were broader than the model, we might expect that the data
represented stars with a range of absolute magnitudes.  Since the data
are narrower than all of the models, no power-law or exponential disk
model is a good to fit the data.  The spheroid models are all very poor
fits to the data.  The only exponential model with any hope of fitting
the data has a scale height of 2 kpc and a scale length of 10 kpc.  Even
this model produces a peak which is a little wide for comfort.

The red line in
Figure~\ref{dwarfthickness}(c) shows the magnitude distribution of all
stars with $117^\circ < \alpha < 130^\circ$, $u^*-g^* > 0.4$, and
$0.44 < g^*-r^* < 0.48$.  The peak in this plot occurs at fainter magnitudes, 
since at the redder colors the stars are intrinsically fainter.  The peak is
narrower because these stars are on the main sequence rather than at
the turnoff, where there is a broader range of intrinsic brightnesses
of stars.  The actual width of the stellar group must correspond to
significantly less than one magnitude in distance modulus.
Figure~\ref{dwarfthickness}(c) also shows a sample model counts of a dwarf
spheroidal galaxy at the distance of the stellar excess, offset from 
the Galactic center, at $(l,b,R_{GC}) = (212^\circ,0^\circ,18\> \rm kpc)$.
The fact that this model nominally fits the SDSS stars counts allows the
possibility of a newly discovered dwarf galaxy in the Galactic plane, though
this is not the only possible interpretation (see \S5.7).

One also verifies that incompleteness at the faint end as a function
of color is not responsible for the turnoff-like feature in
Figure~\ref{cmd125}.  The tests of \S\ref{datared} indicate that this
is not the case for stars with $g^* < 22$, well below the turnoff and
main sequence seen here at $19.4 < g^* < 21.5$.  We adopt the name
S223$+$20$-$19.4 for this structure.

\subsection {Overdensity near $\alpha = 75^\circ$ - S200$-$24$+$19.8}

Now we look at stars at $\alpha = 75^\circ$ in Figure~\ref{Fwedge} and see if those stars can be
explained by smooth components.  See Figure \ref{dwarfthickness2} for
evidence that the excess on this side of the plane is also thinner than
expected for a power-law or exponential disk model.  For this figure,
we tightened up the color range plotted, to reduce contamination from
the thick disk.  As was evident in Figure~\ref{redvsalpha},
the data in this region are of lower quality.  In \S\ref{properties}, we will show
that the measured thick disk turnoff is much redder in this data,
and may indicate a calibration error or incorrect reddening correction
applied here.  Although the absolute photometry is suspect in this
region, we still detect an unexpectedly tight magnitude peak
at $g \sim 19.8$.
Figure~\ref{cmd75} shows a CMD of stars in stripe 82, 
$ 70^\circ < \alpha < 77^\circ$.  This structure is named S200$-$24$-$19.8.

\subsection{Other SDSS data near the Galactic plane - S218$+$22$-$19.5, S183$+$22$-$19.4}

We looked through other SDSS data sets to see if the excess of stars
near the plane showed up in other runs.  We have data at low Galactic
latitude in stripes 12 and 37.  These data include
$153,286$ stars of all colors in stripe 
12 with $122^\circ < \alpha < 135^\circ$ and $252,099$ stars in stripe
37 with $112^\circ < \alpha < 125^\circ$.  Figure~\ref{mondwarfimages} shows wedge plots
for the ends of stripe 12 and 37.

The large increase in stars near the Galactic plane at Galactic latitude
about $+20^\circ$ and $g^* \sim 19.5$ is apparent in stripes 12 and 37 as well.  The magnitudes 
of the stars near the ends of stripe 12 and 37 is similarly as narrow, and peaked
at a similar magnitude, as those
at the end of stripe 10. We adopt the labels S218$+$22$-$19.5 and S183$+$22$-$19.4 for these
apparent overdensities.

Figure~\ref{image37} shows a CMD of stars in stripe 37, separated
by 40 degrees in Galactic longitude from those of Fig~\ref{cmd125}
(at about the same $b$ = +20$^\circ$).  The similarity between
Figure~\ref{image37} and Figure~\ref{cmd125} is
remarkable.  The color-magnitude diagram for S218$+$22$-$19.5 looks the same as well.

\subsection {Fits to the Galactic spheroid}

Now that we have identified several large features in the data which are not
consistent with a smooth distribution of stars, we will attempt to
constrain spheroid models.  In Figure~\ref{dwarfthickness3}, we show the
distribution in magnitude for stars with $0.1 < g^*-r^* < 0.3$, $u^*-g^* < 0.4$,
and $230^\circ < \alpha < 240^\circ$.  We would be very surprised if
these stars were not consistent with a spheroid population.  The figure
shows a very broad distribution in magnitude, most consistent with the
a flattened ($q \sim 0.5$) power-law with slope of $\alpha \sim -3$, though
one could imagine fitting other power-law distributions.  Notice that the
magnitude distribution in this direction is not consistent with an exponential
disk with large scale lengths, especially for $g^*>20$.

We selected all of the stars with $u^*-g^* > 0.4$, $19.0 < g^* < 20.0$, and
$0.1 < g^*-r^* < 0.3$.  The number of these stars as a function of right ascension
is shown in black in Figure~\ref{dwarfboxplot2}.  We then attempted to fit spheroid models
(also shown in Figure~\ref{dwarfboxplot2}) to the data.  We don't expect
many thick disk stars in this plot, since we have selected only
stars bluer than the nominal turnoff.  We expect very few thin disk stars at these
faint magnitudes.  The models are generated
by integrating the models of \S\ref{spheroidmodels} over our apparent magnitude range, 
$19 < g^* < 20$.  With only angular 
information, there is very little difference between models with
different power-law slopes or different assumed absolute magnitudes for the stars.
There is greater sensitivity to the flattening of the spheroid.  The only way to fit
the star counts on both sides of $\alpha = 280^\circ$ is with a very large flattening, such
as $q = 0.5$.  More spherical models can fit the slope near $\alpha = 250^\circ$, but do not
put enough stars at $\alpha = 320^\circ$.  Note that we did not attempt to fit a triaxial
halo, which has been used by other authors \citep{lh96} to explain an interesting 
asymmetry in the blue star counts around the Galactic center.

As a class, the power-law models cannot put enough stars around $\alpha \sim 100^\circ$
to explain the high counts there.  We already encountered difficulty fitting this
feature with a power-law in Figures~\ref{dwarfthickness} and \ref{dwarfthickness2},
but it is good to see the discrepancy in this plot as well.  In order to produce
anything close, one would need a flattening more like $q = 0.2$, which is quite a
bit lower than anyone has previously considered and still doesn't fit well.

Since a power-law spheroid does not seem a good fit to these stars, we look to 
other proposed Galactic components to fit the data.
Evidence for a metal weak thick disk
has been given by \citet{mff90} and \citet{n94}. \citet{cb00}, using kinematics of faint blue
stars, find a scale length for this component of 4.5 kpc.
An exponential disk with a scale height of 2 kpc and a scale length of
10 kpc produces our best fit to the angular data near the Galactic plane.  
We need the large
scale length to put enough stars out at $\alpha \sim 100^\circ$.  This model
gives a surprisingly good fit to the data. It is important to remember
that the stars fit here are not the ones routinely assigned to the
thick disk; they have a bluer turnoff.  This would imply that we are
seeing an `even-thicker-disk' of different metallicity (or age) from the `thick disk.'
The only data that this model contradicts is
the narrow magnitude profiles in Figures~\ref{dwarfthickness} and \ref{dwarfthickness2}.  
It does not explain the fainter star counts in Figure \ref{dwarfthickness3}.  
The center
of the peak near the anticenter is slightly shifted between the exponential
disk model and the data.  This shift cannot be reduced by small changes in the
scale lengths, assumed stellar absolute magnitudes, or the Sun's distance from 
the Galactic center.  It could be reduced by moving the Sun to 0.2 kpc above the
Galactic plane (or by tilting or lowering the extended exponential disk model below the plane of the
thin disk by a similar amount).  We placed the Sun 20 pc above the Galactic plane
in our standard model,
in keeping with recent estimates of this parameter \citep{hgmc95, c95, bgs97, hl95, mv98, cetal01}.

We have proposed two possible explanations for the overdensity S223+20-19.4.  The
first possibility is that it is a previously undiscovered dwarf galaxy (probably
in the process of tidally disrupting), or a stream from a dwarf galaxy.  The other
is that it is part of a smooth, metal-poor Galactic component with a double
exponential profile with about 2 kpc scale height and 10 kpc scale length.  One
cannot produce this many star counts near the Galactic anticenter with a power-law
distribution of stars.  Based on the results of \citet{cetal01}, we do not expect stars with these blue colors in the thick
disk.  We will now show that the star counts do not fit exponential density models
with the previously measured scale heights and scale lengths of the thin and thick
disks.

There are two issues that were not addressed in the previous model fits in 
Figure~\ref{dwarfboxplot2}.  We did not consider that the metallicity of the thick
disk could have changed as a function of scale height.  We also did not use our
knowledge of the local normalization of stars from the various Galactic components
to check whether the stellar distributions could be reasonably attributed to
a known and measured disk component.

To address these issues, we selected all of the stars in a broader color range
($0.2 < g^*-r^* < 0.5$) and in five magnitude ranges: $15.5 < g^* < 16.5$, 
$16.5 < g^* < 17.5$, $17.5 < g^* < 18.5$, $18.5 < g^* < 19.5$, and $19.5 < g^* < 20.0$.
This should contain nearly all turnoff stars in the thick disk and halo populations,
and at the bright end the older turnoff stars in the thin disk.  The densities
of these distributions as a function of right ascension are shown in 
Figure~\ref{five}.  We can now fit models to these data plots as a set.  Since the stars
are redder than in the previous plot, we assume they are fainter, using $M_{g^*} \sim 5.0$,
which is about the absolute magnitude of the Sun.
First, we fit the standard thin disk (scale height 0.25 kpc and scale
length 2.5 kpc) and thick disk (scale height 1.0 kpc and scale length 3.0 kpc, with a
local ratio of turnoff stars of 1:30 of the thin disk stars), shown as a black line in Figure~\ref{five}.  
The models are empirically normalized to the data at only one point.  They are
forced to match the data at $\alpha = 240^\circ$ for stars near $g^* = 17$.  This fit sets
the number of stars in this color range in the thick disk in the solar neighborhood
to a reasonable value.

The star counts for standard thin and thick disk models do not fit.  The discrepancy
is most pronounced for the fainter star counts, where we see too many stars near the
Galactic center and too few stars near the anticenter.  We cannot move stars from the
center to the anticenter by tweaking the assumed absolute magnitudes of the stars, the
distance from the Sun to the center of the Galaxy, or the ratio of thin disk to thick
disk stars.  As we have seen, adding a power-law component will also not help add
star counts near the anticenter.  As we saw in Figure~\ref{dwarfboxplot2},
the way to significantly increase the number of stars at faint magnitudes is to add a
component with larger scale height and scale length.

We could take the thin disk and thick disk models and add an additional exponential disk
model to attempt to fit the data.  Since we now have so many adjustable parameters, we
decided to fit only a `thin disk' and an `metal-weak thick disk (MWTD),' and were able
to fit the data about as well as if standard `thick disk' or `spheroid' components were also included in the mix.  We used a thin disk with a scale height of 300
pc and a scale length of 2.8 kpc, and a MWTD with a scale height of
1.8 kpc, a scale length of 8 kpc, and a thin disk to MWTD ratio of 100:1 in the solar
neighborhood (for stars of this color range).  We also used separate assumed absolute
magnitudes for the two components: $g^* = 5.0$ for the thin disk and $g^* = 4.2$ for
the MWTD.  Note that the brighter stars, assumed to be part of the thinner
disk, have $g^* - r^* \sim 0.25$, while the fainter stars associated with the
MWTD structure have $g^* - r^* \sim 0.40$.  Although one might expect the
absolute magnitude of a more metal poor population to be fainter at the
same color, the color difference (i.e. the thin disk samples stars further down
the main sequence) is more important in this case. The models
are fairly insensitive to our assumed absolute magnitudes.
Figure~\ref{five} shows the model in red, with the thin disk and MWTD components in
green and blue, respectively.  With this model, the brighter star counts are dominated
by the thin disk, and the fainter star counts are dominated by the MWTD.

We do not claim to show from this demonstration that a thick disk is ruled out.  We
have done the exercise of adding the third, thick disk exponential to the model, and it makes
little difference.  If we adjust slightly all of the parameters, a thick disk with reasonable
properties can be easily added to the model.  We refrain from quoting numbers for this,
since there are so many correlated parameters in this model that the individual values
of each parameter may have little meaning.  The thick disk may help adjust the relative numbers
of redder and bluer turnoff stars in detail.  For example, look at the relative number of redder
and bluer turnoff stars at $\alpha \sim 235^\circ$, $g^* = 17$ in Figure~\ref{s10230240cmd}.  Most of
them are the redder population.  Now look the model for $16.5 < r^* < 17.5$ in
Figure~\ref{five}. At $\alpha \sim 235^\circ$, somewhat more than half of the stars are MWTD, or
the bluer population.  We do note that qualitatively in the color-magnitude 
diagrams we see only two distinct turnoff colors, except in Figure~\ref{cmd125} where there is a set
of very bright stars with a very blue turnoff.  We do not see a turnoff that gets steadily
redder with increasing magnitude, or which widely varies as a function of position in the Galaxy
(see, for example, stars with $0.2 < g^*-r^* < 0.5$ in Figure~\ref{s10230240cmd}).

As with our model fits to Figure~\ref{dwarfboxplot2}, the peak near the anticenter is not
well centered on the model fits near $g^* = 18.5$.  If we attempt to adjust our height above the plane to
center the model, then the model becomes a very poor fit at the bright end.  One could
imagine that adding a warp to the MWTD might fix this discrepancy.  Note in Figure~\ref{Fwedge} that
there are multiple apparent overdensities of stars near the Galactic plane at the 
anticenter. At $\alpha \sim 125^\circ$, the overdensities are near $g^* = 15.5$ and $g^* = 19.5$.  
At $\alpha \sim 77^\circ$, the overdensities are near $g^* = 18$ and $g^* = 20$.  We have
no ready explanation for this.

Could all of this be explained without a disrupted dwarf galaxy or MWTD, but by warps or flares 
of the thick (or thin) disk?  A warp in the disk means that the highest
density of disk stars, which is generally in the Galactic plane ($b = 0^\circ$),
shifts a little - to higher or lower Galactic latitude, depending on the direction.
Such a warp has been detected in HI \citep{bt86} 
and possibly in stars as well \citep{cs93,a01}.  The measured
shifts amount to less than half a kiloparsec deviation from the $b = 0^\circ$ plane.
Shifting the thick disk up or down by this amount could throw more or fewer stars into
our dataset at any given location, but would not explain why the stars had a bluer
turnoff than the supposed thick disk stars.

Disk flaring occurs if the scale height of the disk increases with
cylindrical radius from the center of the Galaxy.  This effect has also been
seen in the Milky Way \citep{a01} and perhaps in Andromeda \citep{gcr00}.  But again, 
increasing the scale height with Galactocentric
radius also does not put stars all at the same distance from us, and also does not
explain the bluer turnoff of these stars.
We would expect to see the flare putting stars into our sample at even larger
scale heights at slightly higher Galactic latitudes.
We do not see an excess of stars at $g^* > 20$ and $b > 20^\circ$
near the anti-center.  The excess stops at $g^* \sim 19.5$. 

What we would
need is a thick (or thin) disk which goes out to 14 to 18 kpc from the center of the Galaxy, then
warps sharply perpendicular to the plane, goes up to about $20^\circ$ Galactic
latitude, and then ends abruptly.  This would put stars all at the same distance from us
(since we would be looking straight through the `disk').  The stars in this
`disk' must also have a turnoff with the same color as the spheroid population
of stars in order to match the observations of Figure~\ref{turnoffs}.  
Alternatively, one could construct a
flare model which flares up 18 kpc from the center of the Galaxy, and then decreases
in scale height with distance.

\subsection {S52$-$32$-$20.4}

A look at Figure~\ref{dwarfboxplot2} in directions near the Galactic
center shows that one could not fit the observed density of
stars at $\alpha = 240^\circ$ and $\alpha = 320^\circ$ without
assuming a very flattened spheroid ($q=0.5$).  The exponential model
is also a very flattened structure.  The only way to avoid a
very flattened spheroid is to postulate a large structure around
$\alpha = 300^\circ$ which accounts for the star counts
in this direction.  A look at the distribution of magnitudes in this direction
(Figure~\ref{dwarfthickness4}) shows that the magnitude distribution
does not match our expectations for a power-law spheroid, and thus
there is the possibility of yet more structure at $\alpha=320^\circ$.
We identify this possible structure as
S52$-$32$-$20.4.  


We see below
and in Table 2 that the stars in S52$-$32$-$20.4, which were originally
assumed to be part of the spheroid population as well, appear to have
significantly bluer turnoff stars than those of S6$+$41$-$20.0.  This is a
further indication that at least some of the stars in this direction
are members of another Milky Way structure, and would release us from
the need for a very flattened spheroid population.

If there is a stream at S52$-$32$-$20.4 that is not part of a smooth spheroidal distribution,
then it is possible that we could fit a
rounder $q=0.8$ model, also shown in Figure~\ref{dwarfboxplot2}.

\subsection {Properties of the halo structures\label{properties}}

From positions of the turnoffs in color-magnitude diagrams in the vicinity of the identified structures,
selection criteria were chosen which were intended to favor each
of the structures mentioned above.
The specific selections shown in Figure~\ref{dwarfboxplot}
are:  S223$+$20$-$19.4 and S200$-$24$-$19.8, black, 
($7.05 (g^*-r^*) + 17.24 < g^* < 21$ and $g^*-r^* > 0.1$);
S341$+$57$-$22.5 (Sagittarius), blue, ($21.5 < g^* < 23.5$ and $-0.1 < g^*-r^* < 0.7$);
S167$-$54$-$21.5 (Sagittarius),  red, ($20.5 < g^* < 22.5$ and $0.0 < g^*-r^* < 0.6$);
and S297$+$63$-$20.0, green, ($20.0 < g^* < 21.5$ and $0.1 < g^*-r^* < 0.4$).
The star counts per area as a function of right 
ascension for each of these selections are shown.
Each point on the plot represents the number of stars with the selection
criteria in a region of the sky 2.5 degrees wide in declination by 0.5
degrees wide in right ascension.  The curves were normalized to match near the
Galactic center; the scale factor is indicated in the figure legend.  

We can use this plot to estimate the number of turnoff stars in S167$-$54$-$21.5 (Sagittarius), 
for example. 
The peak of the red curve in the plot is at 1000 stars.  Subtracting off a background of 400 stars,
that leaves 600 stars at the peak.  But the curve has been multiplied by a scale of 2, so
there were only really 300 stars at the peak.  The width of the structure is about 50
degrees, or 100 bins.  Multiplying $0.5 \times 100 \times 300$ for the area of a triangle
gives $15,000$ turnoff stars spread over a $2.5 \times 50 = 125$ square degree area of
sky.

The data curves in Figure~\ref{dwarfboxplot} also show a possible peak in the stellar
density at $\alpha \sim 10^\circ$.  This is faintly evident in Figure~\ref{Fwedge}, but
is not distinguished from an extension of S167$-$54$-$21.5 (Sagittarius).

In addition to counting stars in stellar streams, it is interesting to look for
directional information on the angle at which the streams cross the celestial
equator.  We split the equatorial data into three roughly equal
declination bins: $\delta < -0.4^\circ, -0.4^\circ < \delta < 0.4^\circ,$ and $\delta > 0.4^\circ$.
The star counts in these bins are plotted as a function of right
ascension in Figure~\ref{streamdir}.  The center of S167$-$54$-$21.5 (Sagittarius)
moves from $\alpha = 33^\circ$ to $\alpha = 36^\circ$ when the average declination
goes from $\delta = -0.8^\circ$ to $\delta = 0.8^\circ$.  The slope of this shift,
$\Delta \delta/\Delta \alpha = 1.6/3 = 0.53$ is in excellent agreement
with that predicted for the orbit of a Sagittarius stream at this
position by \citet{ilitq01}, where $\Delta \delta/\Delta \alpha \sim 0.5$.
The northern Sagittarius stream structure, S341$+$57$-$22.5,
appears to move towards lower right ascensions as the declination increases, 
which is the expected sign, though the magnitude of the shift is smaller 
than predicted, suggesting some overlapping of streams may be present here.
The direction of S297$+$63$-$20.0 cannot be distinguished from a track
perpendicular to the equator.  The structures S223$+$20$-$19.4 and S200$-$24$-$19.8 
shift slightly towards lower right ascensions as the declination increases.

The color of the turnoffs of the various overdensities provide an illuminating check on
their identities.  Figure~\ref{turnoffs} shows a the number of stars near the
turnoff as a function of color for stars in the various identified structures.
The structures plotted are: S167$-$54$-$21.5 (Sagittarius), $30^\circ < \alpha < 45^\circ$, $21 < g^* < 21.75$;
S297$+$63$-$20.0, $180^\circ < \alpha < 195^\circ$, $20 < g^* < 20.75$; S223$+$20$+$19.4, $120^\circ < \alpha < 130^\circ,
19.5 < g^* < 20.25$; S167$-$54$-$21.5, $70^\circ < \alpha < 80^\circ$, $19.25 < g^* < 20.0$;
S52$-$32$-$20.4, $320^\circ < \alpha < 330^\circ$, $20.0 < g^* < 20.75$; S183$+$22$-$19.4, $100^\circ < \alpha < 125^\circ$,
$19.5 < g^* < 20.25$, and the stars of the Sagittarius dwarf itself from
Figure~\ref{sagcmd} with $22.25 < g^* < 23.0$.  
The counts for all curves have been normalized to peak at 1400.
We chose the magnitude limits for each
structure to produce the bluest possible turnoff.  
Table 2 lists the colors of the turnoffs of the structures in $g^*-r^*$.

The black line shows the stars which are most likely from the spheroidal population
of stars in the Galactic halo.  (According to our definition, a MWTD would qualify
as a spheroid population.)  It is interesting that the turnoff of the
Sagittarius dwarf, whose photometry has been scaled to match that of S341$+$57$-$22.5
(Sagittarius), is bluer than that of the spheroid population.  Likewise, the turnoff
of S167$-$54$-$21.5 is blue - further evidence that this overdensity is a part of the
tidal stream of the Sagittarius dwarf galaxy.  None of the other identified
structures have turnoffs as blue as these.

S223$+$20$-$19.4, S218$+$22$-$19.5 and S183$+$22$-$19.4 have the same color turnoffs as the spheroid distributions.
However, S200$-$24$-$19.8, which one could imagine might belong to the same halo structure
as S223$+$20$-$19.4 and S183$+$22$-$19.4, has a much bluer turnoff.  We believe the reason for this
can be found in Figure~\ref{tdturnoffs}, which shows the number counts of stars near the
thick disk turnoff ($16.0 < g^* < 16.75$).  Note that the turnoff $g^*-r^*$ color of presumed
thick disk stars in all directions are within a couple of tenths of 0.4, except in the direction of
S200$-$24$-$19.8, which is much bluer than the rest.  (There are very few thick disk
stars in the field of the Sagittarius dwarf itself, which accounts for the apparently poor statistics
of this curve.)  If one adjusted the colors of the S200$-$24$-$19.8
turnoff stars by the amount needed for the thick disk turnoff in this direction
to match all other directions, then the turnoff of S200$-$24$-$19.8 would more closely match the spheroid (S6+41-20.0)
and S223$+$20$-$19.4, S183$+$22$-$19.4.  We are uncertain as to the reason for the discrepancy, but
believe that there is either a calibration error in these data, or the reddening
correction could have been over-applied.  If it is an error in the reddening
correction which produced too blue a color by 0.05 magnitudes, then the $g^*$ and $r^*$
magnitudes should be shifted fainter by 0.15 and 0.10 magnitudes, respectively.

It is interesting that S297$+$63$-$20.0 appears to be intermediate in color between
Sagittarius and the spheroid, as does S52$-$32$-$20.4.  The latter is also a candidate for a flattened
spheroid population.

\section {Discussion}

What have we learned about the halo of the Milky Way from all of this?
The most important lesson is that at distances of 20 kpc from the center of
the Galaxy, the stellar density is not at all smoothly varying,
as a power-law density distribution would be.  It includes
dwarf galaxies, globular clusters, and streamers of tidally stripped stars.
With sufficiently large sky coverage, and good color photometry, these streamers can be identified
by their density in space, and not just by kinematic techniques which
have been previously used to identify moving groups in the solar neighborhood.

The prevalence and ambiguity of clumped stars in our data frustrate our attempts to fit
any smoothly varying `spheroidal' distribution.  The only direction in the sky
in which the stellar distribution looks at all like our expectations for
a power-law distribution is at right ascension $240^\circ < \alpha < 250^\circ$,
where the stars are less than 10 kpc from the Galactic center.  In all other
directions, the stellar distribution appears to be dominated by large
structures with scale lengths of 10 kpc or greater, or by stars that do not fit
neatly into the standard Galactic components.  Even the stars near $\alpha = 245^\circ$
may not be identified as part of a presumed smooth power-law distribution in the halo.

All of our attempts to fit a power-law to the spheroid population of stars, both in magnitude
and right ascension, produced best fits for a very flattened ($q \sim 0.5$)
spheroid.  We do not rule out a $q = 0.8$ spheroid, however, since we cannot
be sure which of our data, if any, represents the spheroid distribution.
We would also like to be clear that even if the spheroid is flattened, that
does not imply a flattened halo dark matter distribution.  One expects that stars
and their associated dark matter are falling into the Galaxy from all directions.

The paper of \citet{ksk94} is a recent work which found evidence for 
a flattened distribution of blue horizontal branch stars amongst
other more spherically distributed populations of objects in the halo.
This effect is discussed in \citet{cb01}, and
references therein, and suggests that a flattened halo dominates 
at $R< 15$ kpc, while the outer halo is spherical.  We imagine that if
one averaged over all of the streams, the distribution could be spherical
at large Galactocentric radii.  Close to the plane the MWTD or a flattened spheroid could
dominate, and the distribution would appear to be flattened.

The overdensities which we have so far identified, and the smooth distributions
of stars which we have assigned to some of the overdensities, do not account
for all of the stars in the dataset.  For example, look at the star counts in
Figure~\ref{dwarfboxplot} at $\alpha \sim 160^\circ$.  There are significantly
more stars here than in any model smooth model fit in Figure~\ref{dwarfboxplot2}.
We could try to
construct a stream profile for S297$+$63$-$20.0 that put stars out this far.  The
spreading of tidal streams in the disruption process is not unexpected,
especially if the mass distribution of the halo is not spherical.
However, in this case the profile would seem contrived to fill in gaps between
the assumed spheroid distribution at $\alpha \sim 250^\circ$ and the
structure at $\alpha \sim 75^\circ$.

There are enormous streams of stars in the halo.  The tidal stream from 
the Sagittarius dwarf galaxy is one of them.  We may have found
additional large streams as described in this paper.  Since our color cut is
relatively blue, we are biased against finding older, or more metal rich
streams with redder turnoff stars.  We are also less sensitive to smaller,
lower stellar density streams.  One might expect that 
there are streams from smaller infalling stellar associations, or 
more disbursed streams from dwarf galaxies which were consumed by
our Galaxy at earlier times in its history.  These smaller or more
disbursed streams might more naturally explain the difference in star 
counts between the models and the data at, for instance, 
$\alpha \sim 160^\circ$.

\subsection { Debris from the Sagittarius dwarf \label{debris}}

We have shown evidence supporting the identification of S341$+$57$-$22.5 and
S167$-$54$-$21.5 with the tidal stream of the Sagittarius dwarf galaxy.  The
color-magnitude diagrams for the stars in these
structures match that of the dwarf itself.  Also, the $g^*-r^*$
color of the turnoff is consistent with that of the Sagittarius dwarf.
We find no reason to doubt the identification of the structures at
$\alpha \sim 210^\circ$ and $\alpha \sim 35^\circ$ as pieces of the
stream of the Sagittarius dwarf galaxy.

One might ask whether any of the other structures identified in this
paper could be part of the Sagittarius stream as well.

We assume from its low stellar density that the structure S297$+$63$-$20.0 
has undergone tidal disruption in the Milky Way.  It is possible that
it could be a part of the tidal stream of the Sagittarius dwarf galaxy.  
Figure 2 of \citet{iils01} shows how debris from the Sagittarius
dwarf is found off the main Sagittarius streamer orbit.  If the F-stars in
S297$+$63$-$20.0 are related to this off-stream debris,
it implies a more disbursed stream than their
$q=0.9$ model predicts. A model closer to $q=0.7$ is needed to explain the
star density relative to that of S341$+$57$-$22.5 (Sagittarius) in terms of
a single tidally precessed stream.  Note, however, that the turnoff
color of the stars in S297$+$63$-$20.0 do not support the idea that they originated
in the Sagittarius dwarf galaxy.  The turnoff color also does not rule out an
identification with the Sagittarius stream; if they are associated it would
imply that the stellar populations changed along a stream.
It is also noteworthy that the S297$+$63$-$20.0 structure lies exactly on the plane of
the Fornax-Leo-Sculptor dwarf galaxies \citep{m94}, though it is much closer to
the Galactic center than any of these dwarfs.

\subsection { The Monoceros -- Canis Major structure \label{moncan}}

The most tantalizing structures we have identified are S223$+$20$-$19.4, S218$+$22$-$19.5 and S183$+$22$-$19.4, which
may be two sides of the same contiguous structure.  S200$-$24$-$19.8 could also belong to
this structure, but its relationship is more difficult to establish, due
to the lower data quality in this region.
Though it is possible that we have found three independent, similar structures
of stars in the halo, we find that possibility unlikely.
In \S5.7 we explored the possibility that this structure was part of a metal-weak
thick disk.  In this section, we explore the possibility that it is a tidally
disrupted dwarf galaxy spread
across $45^\circ$ on the sky and 11 kpc from us (18 kpc from the center of
the Galaxy).  We believe this would not have been identified
previously because it is so large and close, and it is hidden by
the plane of the Milky Way.  Figure~\ref{dwarfcenter} shows our knowledge
of the edge of this stellar structure. 

Since we do not probe the full extent of any structure in this area
of the sky due to the intervening Galactic plane, it is difficult to
distinguish a disrupted galaxy residual stream from a dwarf galaxy.  Without kinematic
information it is difficult for us to identify streams with possible
parent dwarf galaxies.  Since we do not see much of the perimeter
of this structure, we cannot distinguish very easily between a dwarf
galaxy and a gravitationally unbound
streamer which circles the entire Galaxy at an inclination $i < 20^\circ$
to the Galactic plane (or something in between).  Most other orbital
directions are ruled out because it is only evident at the ends of
stripes 10, 11, 12, 37, and 82. 

A dwarf galaxy or disrupted galaxy stream of stars in the Galactic halo provides a
simpler model which produces stars all at about the same distance from us.
With a distance modulus of about 15.2 (from Figure~\ref{dwarfthickness}
and an assumed absolute magnitude of $g^* = 4.2$), the distance to S223$+$20$-$19.4
is 11 kpc from the Sun.  The same distance is derived for S183$+$22$-$19.4.  These two
overdensities are separated by $40^\circ$ on the sky.  If they indeed belong
to the same structure, the structure is at least 8 kpc across.  This is of the
same scale as other large structures identified in the halo, including
the Sagittarius dwarf spheroidal galaxy and the Sagittarius dwarf streamer.  
If S200$-$24$-$19.8 is part of the same structure, it is 8 kpc in the declination
direction.  

The width of the main sequence of S223$+$20$-$19.4 (see Figure~\ref{dwarfthickness}) is
only about one magnitude wide at $g^* = 21.1$.  From the errors in color alone
(multiply the expected dispersion in $g^*-r^*$ color at $g^* = 21.1$, from Figure~\ref{limmag},
by the slope of the main sequence in Figure~\ref{s10230240cmd}),
we could explain this entire width.  To gain an upper limit on the thickness
of the structure, we assume the entire one magnitude dispersion is due
to depth of the structure, and obtain an upper limit for the depth of the
structure of 6 kpc.  We obtain a similar measurement for the depth
of S183$+$22$-$19.4.

We now ask what the mass of a satellite in the Galactic plane would have to be in
order to remain tidally bound.  A simple tidal analysis can be done
following, for example, equation 7.84 in \citet{bt87}.  The mass of 
the satellite within the tidal radius is given by:
\[m_{sat} = 3 M_{MW} (\frac{r_{tidal}}{D})^3,\]
where $r_{tidal}$ is the tidal radius of the dwarf, $D$ is the distance of the
satellite from the center of the Milky Way, and $M_{MW}$ is the mass of the
Milky Way within a radius of $D$.  This equation holds for $m_{sat} << M_{MW}$
and $r_{tidal} << D$.  
Plugging in $r_{tidal} = 4$ kpc and $D = 18$ kpc,
we find that the satellite would have to have a mass equal to $3\%$ of the
mass of the Milky Way.  Estimating the mass of the Milky Way interior to $D$
from $M_{MW} = v_{MW}^2 D / G$ with $v_{MW} = 220$ km/sec, we find $M_{MW} = 2 \times 10^{11}$ M$_{\odot}$, and an inferred satellite mass of $6 \times 10^9$ M$_\odot$.

For reference, the dynamically estimated initial mass of the Sagittarius dwarf galaxy 
is between $10^9$ and
$10^{11}$ M$_\odot$ \citep{il98, jb00}, and it is currently about $10^9 M_\odot$ \citep{jmsrk99}.  Sagittarius is
located 16 kpc from the Galactic center, and
prolate with axis ratios 3:1:1 and a major axis of
at least 9 kpc \citep{iwgis97}.  So far, our observations could be explained by
a dwarf galaxy, similar in size to the Sagittarius galaxy, and hiding in the plane of the Milky Way
18 kpc from the Galactic center.  

We now ask whether the star counts support the existence of so massive a structure in the halo.
For stars of this turnoff color, $g^*-r^* = 0.28$, a relatively metal-poor, spheroidal type
population with $[Fe/H] = -1.7\pm 0.3$ is implied.  An isochrone analysis
like that for the Sagittarius stream of \S\ref{ghostsag} then indicates
that these turnoff stars typically would have masses near $0.75 M_\odot$ and approximate
ages of 13 Gyr.

It is difficult to estimate the total number of stars in the proposed structure.  If it
is a dwarf galaxy, we have only detected the tails of the distribution.  Star counts must
be estimated by extrapolation.  The highest detected stellar density is about 1500
F and G stars above background in a 1.25 square degree region of the sky (Figure~\ref{dwarfboxplot}).  A structure
with constant stellar density over a $40^\circ \times 40^\circ$ area of the sky would contain
$2 \times 10^6$ stars.  This is a lower limit.  

If, instead, one fits to a model power-law distribution ($\alpha = -3.5$) of 
stars centered half way between our two detections, such as that shown 
in Figure~\ref{dwarfthickness}c and Figure~\ref{dwarfboxplot}, we 
calculate $1 \times 10^7$ F and G stars in the whole structure.  If we put 
the center of the dwarf galaxy in the plane of the Milky
Way, the inferred star count is several times higher.  One could increase or decrease
the inferred mass in stars by suitably adjusting the axial ratios or density profiles
of the models.  A mass in stars of a few times $10^8$ solar masses is feasible, though 
by no means proven.

The total number of stars in the structure could easily be  larger than these
estimates if it is part of a stream which circles the Galaxy.
If the stream contains
at least $2 \times 10^6$ stars in the $40^\circ$ section of sky where we detected
it, and if it extends all the way around the Galaxy with similar density, it must contain
at least $2 \times 10^7$ stars.  The actual stellar and dynamical masses are likely
to be much higher, since these estimates use the lowest possible extent and stellar
densities.

Thus, the overdensity could indicate a dwarf galaxy in the constellation Monoceros or 
in nearby Canis Major to the South.  However, even if it is a dwarf galaxy, the high 
tidal mass calculated
above suggests that it would be in the process of disrupting,
just as the Sagittarius dwarf galaxy is.  One could go a step further, and suggest that
what we have detected is not a dwarf galaxy at all, but is instead a gravitationally
unbound stream of stars.  This conclusion might be preferred, since it frees
us from explaining the coincidence of having found the very ends of 
the structure by chance in stripes 10, 82 and 37.  Figure~\ref{dwarfboxplot}
shows that even as a stream, this structure is significantly denser than
the Sagittarius stream where it crosses zero declination.  It is not only denser where we
detect it, but it is also steeply rising as we run out of data.  

If it is the result of the
complete disruption of a gravitationally bound group of stars, the original mass of the
infalling matter was probably quite large.  The stream must contain
at least $1 \times 10^6$ stars in the $40^\circ$ section of sky where we detected
it.  If it extends all the way around the Galaxy with similar density, it must contain
at least $1 \times 10^7$ stars.  The actual stellar and dynamical masses are likely
to be much higher, since these estimates use the lowest possible extent and stellar
densities.

%
\section {Conclusions}

From stars in the Sloan Digital Sky Survey, we have shown that we can detect large ($\sim 10$ kpc)
structures of stars in the halo of the Milky Way.
In Paper I, we showed that substructure in the Galactic halo could be identified
from photometric data for blue stars.  In this paper we extended the technique 
to identify large structures directly from turnoff
stars.  The color-magnitude diagrams of the stars in the structures should resemble
the color-magnitude diagrams of the original dwarf galaxies or clusters which
fell into the Milky Way.  Features of the diagrams can be used to constrain the
origins of each detected overdensity.  

As more data are collected from the Sloan
Digital Sky Survey, we will be able to trace each structure through space, and connect
the overdensities in each stripe to each other to build up a large scale map of
large stellar streams in the halo of our Galaxy.  For now, we must be content to
identify and name each overdensity, and only to estimate their full extent and
origin.  In this paper, we studied the $g^*-r^*$ colors and $g^*$ magnitude distributions 
of seven overdensities of halo stars in the equatorial plane.  We also show overdensities
in three off-equatorial stripes, since they appear to be associated with
the equatorial structures.

We emphasize these conclusions:

1. We show additional evidence that the overdensities S341$+$57$-$22.5 and S167$-$54$-$21.5 are in
fact part of the tidal stream of the Sagittarius dwarf galaxy.  These structures
were discovered Paper I, and were interpreted by \citet{iils01} as two slices
through the tidal stream of the Sagittarius dwarf galaxy.

The color-magnitude diagram of S341$+$57$-$22.5 bears striking resemblance to the color-magnitude
diagram of the Sagittarius dwarf, including similar clumps of red stars.  The
color-magnitude diagram of S167$-$54$-$21.5 is consistent with that of the Sagittarius dwarf
galaxy.  In addition, the two overdensities are shown to have the same color turnoff
stars as the Sagittarius dwarf galaxy; the turnoff of the Sagittarius dwarf is 0.08 to 0.1
magnitudes bluer in $g^*-r^*$ than the assumed Galactic spheroid stars, and 
substantially bluer than any other structure we have identified.

A comparison
of the number of stars detected in the clump of red stars of S341$+$57$-$22.5 and
S344$+$58$-$22.5 (in the adjacent stripe 11)
with the number of similar stars in the Sagittarius dwarf indicate that we see about
$1.5\%$ of the present stellar mass of the Sagittarius dwarf in this 110 square degree 
area of the sky.  This result assumes a constant clump star to stellar mass ratio between
the Sagittarius dwarf and the stream.

2. From the spatial and  magnitude distribution of turnoff stars 
in the spheroid, there is clear evidence for a diffuse 
structure S297$+$63$-$20.0, at a distance of about 20 kpc, extending over tens 
of degrees.
Other evidence for this structure is a possible group of clustered RR Lyraes
noted at the same distance and position by \citet{v01}. 
This structure is very close in position to the
Sagittarius dwarf tidal stream at S341$+$57$-$22.5, but two magnitudes brighter.  

Although its proximity
to the Sagittarius stream suggests that it might be another part of this same disrupted
galaxy, 
the color of its turnoff ($g^*-r^* = 0.26$) is not the same.  It is
intermediate between that of the Sgr dwarf ($g^*-r^*=0.22$) and that 
of the spheroid ($g^*-r^* = 0.28$). 
Surprisingly,
the turnoff of S297$+$63$-$20.0 is nearly the same color as the turnoff of S52$-$32$-$20.4.

3. We observe many more stars at low Galactic latitudes near the Galactic
anticenter than standard models predict at $g^* \sim 19.5$.  These stars were selected to be
bluer than the turnoff of the thick disk stars.  
Several of our identified structures lie in this general direction, and
may be part of the same physical structure in the Galaxy.
The structures S223$+$20$-$19.4, S218$+$22$-$19.5, S183$+$22$-$19.4, and with less significance S200$-$24$-$19.8, have similar color-magnitude
diagrams, turnoff colors, and inferred distances.  The similarity between the color-magnitude
diagrams for S223$+$20$-$19.4 and S183$+$22$-$19.4 is particularly striking.  The narrow main sequence
seen in the color-magnitude diagrams is consistent with stars all at the same distance,
about 11 kpc from the Sun.  
S223$+$20$-$19.4 and S183$+$22$-$19.4 are separated by $40^\circ$ in
right ascension.  These are both separated from S200$-$24$-$19.8 by $40^\circ$ in declination.
The inferred spatial extent of the structure is 8 kpc in declination, centered approximately
on the Galactic plane, by at least 8 kpc in right ascension.  Since it would seem
coincidental to have detected the structure exactly at its ends, we expect the structure
is substantially longer than 8 kpc.  The inferred distance from magnitudes of turnoff
stars is 11-16 kpc from the Sun.
From the magnitude distribution, the structure is less than 6 kpc thick along the line of sight.  
The turnoff stars of this structure have colors
of spheroid stars ($g^*-r^* = 0.28$), rather than
colors of thick disk stars ($g^*-r^* = 0.40$).  

We propose two possible explanations for the unexpectedly high concentrations of blue stars
near the Galactic anticenter.  One of the possibilities is that they are stars associated with
a tidally disrupted dwarf galaxy.  The other is that these stars are part of an `even thicker
disk' population which has a bluer turnoff than the thick disk, a scale height of about 2 kpc,
and a scale length around 10 kpc.  Though neither explanation explains all of the data, either model
could be reasonably extended to work.  We do not propose that these possibilities exclude all other models - 
they are merely the most reasonable explanations we could find.

The tidally disrupted dwarf galaxy model neatly explains a distribution of stars all at the
same apparent distance.  The presence of the disrupted Sagittarius dwarf galaxy proves that
such structures can and do exist in the Milky Way halo, and that they can be detected by these
techniques.  The inferred physical parameters for such a structure, though large, are not 
prohibitive; the projected mass of the original dwarf galaxy could be
of similar size to the Sagittarius dwarf galaxy.
This model does not explain why the stars we see towards the Galactic
center show an unexpectedly large flattening (other additional streams or Galactic
components are required to explain this), or why the turnoff of this proposed dwarf
has the same color as the stars towards the Galactic center.

The `even thicker' double exponential disk model 
uses large scale lengths to put the peak of the stellar density at faint
enough magnitudes.  This model is appealing because it naturally explains why such a structure
is found over at least $40^{\circ}$ of right ascension in the Galactic plane, and may correspond
to the `metal-weak thick disk' proposed by previous authors.  The negatives of this model are
that it does not fit the faint star counts near the Galactic center (Figure~\ref{dwarfthickness3}),
and consumes all of the stars brighter than 20th magnitude which we expected were part of the
power-law spheroid part of the halo.  It also is rather broader in magnitude, spreading stars over a larger distance range, than the data suggest.  This model would reduce the significance of, or
eliminate, a power-law distribution of halo stars.  The distribution in magnitude of the concentration
of stars near the anticenter is somewhat narrower than expected for an exponential disk
(Figure~\ref{dwarfthickness}).

Neither of these models explains the excess of stars at $15^{th}$ and $17^{th}$ magnitude near
the plane at the Galactic anticenter.  A stream model might introduce additional streams to 
explain this, whereas a disk model might introduce warping to explain this.

%


4. On the other side of the Milky Way at $(l,b) = (52^\circ,-32^\circ)$,
in a direction not far from the Galactic Center S52$-$32$-$20.4, there is 
evidence for stars distinct from
a smooth spheroidal distribution of stars at $g^* = 20.8$.  The distribution
in magnitude is not consistent with a power-law spheroid, although
it could be fit with an exponential disk with large scale length.  This
is in contrast to the structure S6$+$41$-$20.0, which is the only observed
concentration of halo stars that shows the spatial distribution, both
in right ascension and apparent magnitude, expected for a power-law Galactic
spheroid.  Additionally, the turnoff color of the stars in S52$-$32$-$20.4 is not
the same as the presumed spheroid stars at S6$+$41$-$20.0, but rather intermediate
between the spheroid and the Sagittarius dwarf.

If we try to fit a power-law to both the stars in S6$+$41$-$20.0 and S52$-$32$-$20.4,
then we must have $q < 0.6$ (and there is a poor fit with distance in
the direction of S52$-$32$-$20.4).
If this structure is regarded as distinct from
the spheroid, then the remaining spheroidal stars towards the
Galactic center could be fit with a rounder model, $q=0.8$.

We do not present a single, coherent proposal for the components of the Milky Way, 
since it is not clear to what component each identified overdensity should be
assigned.  As more SDSS data is analyzed, and the extent of each structure
is better known, we hope to generate a more coherent, defendable model.

5.  Aside from the obvious large overdensities in the halo, there is tantalizing evidence for further,
smaller structures, for example at at $\alpha = 10^\circ$.  One could imagine that there are
even smaller structures which are not spatially resolved, which make up the difference between
the observed star counts and the model fits to the spheroid population.

In this paper and in Paper I we identified seven or eight large overdensities which we believe
might be associated with three or more halo structures.  In view of these results,
one must take seriously the possibility that there are many such previously unidentified
structures in the halo.  It is also probable that there are many smaller or more
disrupted structures which might be better detected from kinematics than spatial 
information.
Models of structure formation which have produced ``too many galaxies per halo" may actually
be predicting correct numbers of smaller halo structures.  It appears we 
may be able to solve the problem by observationally finding more 
disrupted satellites in each halo.

One cannot help but wonder many things about the results presented in this paper.  We conclude 
our discoveries with a list of questions for which we do not yet have answers.  Is there
a previously undiscovered dwarf galaxy hidden in the plane of the Milky Way?  If there is a massive streamer
or dwarf galaxy which orbits our Milky Way in the Galactic plane, could this disrupt the
disk at about 18 kpc from the Galactic center?  Would it cause disk warping or flaring?
Are there any dynamical models that could put a sheet of stars in a ring around the Galaxy
without an infalling dwarf?  If there is a structure with a scale length of many kiloparsecs
which is only 11 kpc from us, can we detect any stars from this structure in the solar
neighborhood kinematically or photometrically?  Is there a metal-weak thick disk?
Is there a model for the metal-weak thick disk which could explain why the stars seem to
be shifted towards lower Galactic latitudes at $g^* \sim 18$?
Why don't we see a break, or at least a gradient, in the turnoff color between stars
which would nominally be assigned to the thin disk and those which would be assigned to
the thick disk? Finally, are there any halo stars which form a well-mixed, smooth, spheroidal
distribution, and were any of them formed during the initial collapse of our Galaxy,
as was proposed by \citet{els62}?

\acknowledgments

We acknowledge useful comments from Hugh Harris, Daniel Eisenstein, 
Amina Helmi, Doug Whittet and David Weinberg.   We appreciate the close reading
and extensive comments of the anonymous referee, which substantially improved
the paper.

The Sloan Digital Sky Survey (SDSS) is a joint project of The
University of Chicago, Fermilab, the Institute for Advanced Study, the
Japan Participation Group, The Johns Hopkins University, the
Max-Planck-Institute for Astronomy (MPIA), the Max-Planck-Institute
for Astrophysics (MPA), New Mexico State University, Princeton
University, the United States Naval Observatory, and the University of
Washington.  Apache Point Observatory, site of the SDSS telescopes, is
operated by the Astrophysical Research Consortium (ARC).

Funding for the project has been provided by the Alfred P. Sloan
Foundation, the SDSS member institutions, the National Aeronautics and
Space Administration, the National Science Foundation, the
U.S. Department of Energy, the Japanese Monbukagakusho, and the Max
Planck Society. The SDSS Web site is http://www.sdss.org.

\clearpage


\figcaption {Two dimensional ($g^*$ and RA) polar coordinate-density 
histogram of turn-off stars on the celestial 
equator with $0.1 < g^*-r^* < 0.3$.  The shading of each cell indicates
the relative number counts of stars in each (RA, $g^*$) bin.  Typical
absolute magnitudes of stars with these colors are $M_{g^*} = +4.2$,
and thus stars with $g^* = 19.4$ are at distances of 11 kpc from
the Sun. $g^* = 22.5$ corresponds to objects 45 kpc from the Sun.
The center of the Galaxy ($l=0$) is towards the lower left 
at $\alpha = 228^\circ$.  The intersection of the plane of the Celestial
equator with the Galactic plane ($b=0$) is indicated by the bold, black line.
Note the numerous high signal-to-noise structures existing in 
the halo of the Milky Way.  Boldface labels indicate positions of
overdensities which are discussed in the text.  The color cut excludes most thick disk
stars.  The feature at $\alpha = 60^\circ$ is probably an artifact of the reddening
correction applied to the data,
since there is a large dust cloud at this position.  The streak at
$\alpha = 229^\circ$ is due to Pal 5.  The grey scale bar at the bottom
of the figure indicates relative star count density in each pixel.
\label{Fwedge}}

\figcaption {Reddening and quasar candidate counts vs. RA on the celestial 
equator ($-1.26^\circ  <\delta < 1.26^\circ$).  E(B-V) at $\delta = 0$ (scaled $\times 10$)
is shown to indicate areas where dust may affect the object selection.
The quasar candidates are all the stellar
objects in the color box 
$18 < g^* < 21.5, 0 < u^*-g^* < 0.3, 0.1 < g^*-r^* < 0.3$.  The fact 
that the quasar candidate counts are generally uniform around the
equator indicates that selection of F stars as a function of $\alpha$ is
also uniform.
\label{redvsalpha}}


\figcaption {Limiting magnitude and completeness for three color ranges, and also
$g^* - r^*$ errors vs. color.  Stars in overlapping runs were matched based on
position (within $2''$).  Then stars in the first run with colors in the 
three ranges indicated were examined for matches (of any color) in the other run as 
a function of magnitude.  The plotted results (with
binomial distribution error bars shown) indicate completeness independent of
color to about $g^* \sim 22.5$.  Shown as crosses are rms errors in
$g^*-r^*$ for objects with $g^*-r^* < 1$ as determined from matching objects.
\label{limmag}}


\figcaption {Color-Magnitude diagram of objects typed as galaxies in
the SDSS sample. 
Normally, color-magnitude diagrams are shown with one dot plotted for
each star observed.  Since we have too many stars to be effectively
plotted in this way, we instead histogram the number
of counts in each 2-D binned cell.  The horizontal bin width in 
$g^*-r^*$ color is 0.02 magnitudes,
and the vertical bin width in $g^*$ is 0.05 magnitudes.  The image
is plotted with square root density scaling.
The galaxies are localized in color magnitude space redward of the
turnoff in such a way that they do not contaminate blue star counts
at $g^* < 22.5$.  \label{galaxies}}

\figcaption {Color-magnitude image of S341$+$57$-$22.5 (Sagittarius).
This shows stars in the equatorial region $200^\circ < \alpha < 225^\circ$,
with binning in color and magnitude identical to Figure~\ref{galaxies}.
Capturing the color-magnitude diagram as an image allowed us to subtract off
images of other parts of the sky which do not contain the Sagittarius dwarf
streamer, as detailed in the text.  
The image clearly shows a turnoff at about $g^* = 22.5$, a giant branch extending
to a clump of red stars at $g^* = 19.7$, a blue horizontal
branch extending from the clump to as blue as $g^*-r^* = -0.1$, and 
blue stragglers which extend from the turnoff up through the horizontal
branch.  
\label{sagsubn}}


\figcaption {Color-magnitude image of Sagittarius dwarf.  The stellar
data used to create this color-magnitude image comes from
\citet{mbcimpp98}, who published photometry for two fields in the
Sagittarius dwarf itself, using V and I filters on the 3.5 m ESO-NTT
telescope.  
In order to compare this data with Figure~\ref{sagsubn}, the photometry was 
converted to the SDSS filter system and the magnitude shifted to reflect
the difference in distance between the Sagittarius dwarf stream (at this
position) and the Sagittarius dwarf galaxy.  The
similarity between the color-magnitude diagram of the Sagittarius
dwarf and that of the streamer in Figure~\ref{sagsubn} is striking.
\label{sagcmd}}


\figcaption {Color-magnitude image of S167$-$54$-$21.5 (Sagittarius).
We used data from stripe 82, $15^\circ < \alpha < 50^\circ$ to make a
color-magnitude image, and then subtracted suitable reference fields
to decrease the contrast between the streamer stars and the other
stars of the Galaxy.  The southern streamer shows a clear turnoff,
giant branch, and blue straggler stars.  The horizontal branch at $g^*
\sim 18$ is rather weak, and the clump of red stars is neither ruled out nor apparent.
The color-magnitude diagram is consistent with
its identification as a piece of the Sagittarius streamer, though we
cannot make a positive identification from this diagram.
\label{sagsubs}}


\figcaption {Color-magnitude image of stars in 
stripe 10 with $230^\circ < \alpha < 240^\circ$ (S6$+$41$-$20.0, spheroid).  
\label{s10230240cmd}}


\figcaption {Excess stars in S167$-$54$-$21.5 (Sagittarius) as
a function of apparent $g^*$ magnitude.  This shows all of the stars
in stripe 82 with $0.1 < g^*-r^* < 0.3$ and $30^\circ < \alpha < 45^\circ$, with
all of the stars in $20^\circ < \alpha < 25^\circ$ and $45^\circ < \alpha < 55^\circ$
subtracted off.  We use the apparent magnitude of the turnoff stars from this
plot plus the distance modulus to S167$-$54$-$21.5, as determined from horizontal
branch stars, to estimate that the turnoff stars have $M_{g^*} \sim +4.2$.
\label{Fstarmag}}


\figcaption {Models of power-law and exponential disk profiles as they
intersect the celestial equator.  Four smooth models are shown.
a), b) and c) are power-law models with $\alpha=-3.5$ and flattening q=(0.5, 1.0, 1.5),
respectively.
d) shows an exponential disk model with a scale length of 3 kpc and a scale
height of 1 kpc.  All models assume a narrow (delta-function)
population of F main sequence stars of absolute magnitude $M_{g^*} = 4.0$ for
power-law models and $M_{g^*} = 4.5$ for the exponential disk model.
These absolute magnitudes were chosen to put stars at about the right
apparent magnitude near the anticenter, and are not too far off from
the absolute magnitude we expect for stars in Figure \ref{Fwedge}.
\label{models}}

\figcaption {Hess diagram for stars
in stripe 10 with $180^\circ < \alpha < 195^\circ$ (Spheroid).
\label{image-S10-180-195}}

\figcaption {Hess diagram for stars with $117^\circ < \alpha
< 130^\circ$ S223$+$20$-$19.4.  Note the turnoff near $g^* \sim 19.4$ which is distinct
from the brighter thin/thick disk stars at $g^* < 18.5$.  This structure appears
to be the main sequence of a localized (in distance) distribution of stars about
11 kpc from the Sun in the direction of (l,b) = ($220^\circ,20^\circ$).
\label{cmd125}}

\figcaption {Model fits to the data at $122^\circ < \alpha < 130^\circ$ (S223$+$20$-$19.4). 
In a) six power law models are plotted. In b) six exponential disk
models are plotted.  Models and data are arbitrarily normalized.  Note how broad in magnitude these
distributions are compared to the peak of the distribution of observed stars in c).
None of the models of a) or b) fit the data well, though an exponential disk model
with scale height $\sim 2$ kpc and scale length $\sim 10$ kpc
might be made to work by postulating brighter and fainter peaks in the star counts from other
populations.  In c) we also show a distribution
of redder stars.  The redder color cut includes thick disk stars at the bright end,
and an even narrower peak (the peak is fainter because the selected stars are
intrinsically fainter).  A model spheroid offset
from the center of the Galaxy by about 18 kpc in the direction of the excess stars is
fit to the data.  One expects the broader peak in the data than in the model due to
intrinsic spread in the absolute magnitudes of the stars, plus photometric error.
\label{dwarfthickness}}

\figcaption {Model fits to the data at $70^\circ < \alpha < 77^\circ$ (S200$-$24$-$19.8).  Same
as Figure~\ref{dwarfthickness}, except on the other side of the Galactic
plane.  Again, the power-law models look nothing like the data.  Exponential
disk models are better, but have a somewhat broader distribution and require
larger scale lengths than typically assumed for the thick disk.
\label{dwarfthickness2}}

\figcaption {Hess diagram for
stars on the end of stripe 82 with $70^\circ < \alpha < 77^\circ$ (S200$-$24$-$19.8).  
\label{cmd75}}

\figcaption {These 2D histograms of F star density
are made in the same way as those in Figure~\ref{Fwedge}.  Data from stripes
37 (upper), and 12 (lower) are shown.  
An excess of stars at $g^* \approx 19.4, b \sim +20^\circ$ is seen
in each plot.  These stars all have color-magnitude
diagrams similar to that of Figure~\ref{cmd125}, and may be pieces of
a vast structure (dwarf companion or large scale length
exponential disk).
 \label{mondwarfimages}}

\figcaption {Hess diagram for
stars on the end of stripe 37, with $\alpha < 125^\circ$ (S183$+$22$-$19.4).  Note the similarity to
the Fig~\ref{cmd125}.  The horizontal branch at $g^* \sim 20.5, g^*-r^* \sim -0.2$ is from the globular cluster
NGC 2419.
\label{image37}}

\figcaption {Model fits to the data at $230^\circ < \alpha < 240^\circ$ (Spheroid). 
Here the flattened power-law model $q=0.5$ is a good fit to the data (the normalization
of the models is arbitrary).
Note in particular that no exponential disk model is a good fit.  If we wish
to explain the structure in Figure~\ref{dwarfthickness}c with a smooth,
exponential disk distribution of spheroid stars, one must modify the functional
form of the spheroid or add a separate component in the general direction of the Galactic center.
\label{dwarfthickness3}}

\figcaption {F star counts along the celestial equator.
We show number counts of stars in stripe 10 and 82 with 
$20.0 < g^* < 21.5$ and $0.1 < g^*-r^* < 0.4$ (black).  The data are binned
in half degree bins in right ascension.  Since the stripe is 2.5 degrees
wide, the number counts are per 1.25 square degrees. 
We also show models for exponential disk and power-laws, scaled to best match
the data.  None of the power-law models put large numbers of stars near the
Galactic anticenter.  It is possible to put stars near the plane and the
anticenter with an exponential disk model.
Also a prolate spheroid centered out at (X, Y, Z) = (-17.4, -4.3, 1.1), which
could represent a dwarf spheroidal galaxy, is 
shown in magenta.  This is a very good fit to the data near $\alpha = 100^\circ$.
We can than then fit the stars near $\alpha = 280^\circ$ with a power
law model, but only if it is very oblate.  If we wanted to represent the spheroid by
a more spherical power law, then the stars near $\alpha = 320^\circ$ would need to be explained
by a different Galactic component.
\label{dwarfboxplot2}}

\figcaption {Thin disk plus extended metal weak thick disk model of the Galaxy.  We show
F/G turnoff star counts as a function of right ascension, on the Celestial Equator, for five 
magnitude ranges.  The stars were color selected, with $0.2 < g^*-r^* < 0.5$.
  The data are indicated by black dots (which show up as a thick black line at
zero counts where there is no data).  The solid lines show theoretical
exponential disks fit to the data.  The standard (black) and proposed (red) thin+thick
exponential disk models are normalized to match the star counts at $\alpha = 240^{\circ}$
in the $16.5 < g^* < 17.5$ plot.  A `standard model' for the thick and thin disk,
with thin disk scale height 250 pc, thin disk scale length 2.5 kpc, thick disk scale 
height 1.0 kpc and thick disk scale length 3.0 kpc, is shown in black.  The assumed
ratio of thin disk stars to thick disk stars at the solar position is 1:30.  This model
does not account for the large number of faint stars near the Galactic anticenter.
Adding a power law spheroid to this would only make the disagreement larger.
In red we show a model with a thin disk with scale height 200 pc and scale length 2.8 kpc,
plus a proposed metal weak thick disk with scale height 1.8 kpc and scale length 8 kpc.
The ratio of thin disk to MWTD stars at the solar position is 100:1.  The thin disk
and MWTD components of this model are shown in green and blue, respectively.  One
could achieve a similar fit to the data if a third double exponential, representing
a standard thick disk, is included in the model.  It is difficult to fit the peaks near
the center and anticenter by including both standard power law and MWTD components.  The
power law inserts too many stars near the Galactic center (and very few towards the
Galactic anticenter) at faint magnitudes.  The heavy black line at zero counts
indicates regions where no data is present.
\label{five}}

\figcaption {Disk and halo models vs. data at $320^\circ < \alpha < 330^\circ$ (S52$-$32$-$20.4). 
Note that this distribution is not consistent with a power-law spheroid.  To fit
this distribution with an exponential disk would require large scale lengths.
\label{dwarfthickness4}}

\figcaption {Selected F star counts along the celestial equator.
We show the relative distribution of stellar densities along the
celestial equator.  Four boxes in color-magnitude space were chosen to highlight stars
in each of the detected overdensities (S167$-$54$-$21.5 (Sagittarius) red; S223$+$20$-$19.4 black; S297$+$63$-$20.0 green; and S341$+$57$-$22.5
(Sagittarius) blue.   Curves were scaled so that they match
in height at $\alpha = 240^\circ$.
Notice that each detected overdensity produces a strong peak in this plot.  There is
an additional overdensity apparent in this plot at $\alpha = 10^\circ$.  One could imagine
other smaller overdensities that could help fill in the plot between the theoretical
curves and the data curves.  The dip in star counts in the center of the structure at
$\alpha=213^\circ$ is artificially caused by a decrease in the data quality in this region.
The very low points occur where the data quality was so low that the data in close
to half a degree of one of the runs was removed.
\label{dwarfboxplot}}

\figcaption {Stream direction.  The SDSS data are obtained in stripes of width 2.5 degrees
in declination.  To examine the direction
that a stream of stars makes with a stripe, we divide each stripe up into
three sub-stripes, each with width $0.8^\circ$ in declination. 
The histogram of star counts within each sub-stripe is shown for a) S167$-$54$-$21.5 (Sagittarius),
b) S341$+$57$-$22.5 (Sagittarius), c) S297$+$63$-$20.0 and d) S223$+$20$-$19.4.  Each plot shows the stars which
have colors and magnitudes consistent with the clumped population (see \S\ref{properties}).
For the S167$-$54$-$21.5 (Sagittarius)
structure a shift is apparent from south to north and indicates that the 
stream crosses the celestial equator at an angle of PA$\sim 30^\circ$.  A shift in the
opposite direction (as would be expected for the stream) is observed in S341$+$57$-$22.5.
The peak at $\alpha = 229^\circ$ is due to the globular cluster Pal 5.
In this figure, the counts in a single three degree bin have been divided by 0.925 to compensate for
missing data in right ascension range $215.5^\circ < \alpha < 215.95^\circ$.  This data 
was flagged as bad data and was
removed from our dataset.  Data near this right ascension has poorer than average seeing.
The density of stars in S223$+$20$-$19.4 decreases for increasing declination.
Since we do not detect the a peak for structure S223$+$20$-$19.4, we cannot tell whether this is a
shift or a density change in the declination direction.  If it is a shift, it is consistent
with a structure which is aligned more along the Galactic plane than along the celestial
equator.
\label{streamdir}}

\figcaption {Color cuts of stars in the vicinity of
the turnoff for several populations (isolated in $\alpha$ and $g^*$ magnitude) of stars.
The peak of the histogram for each population gives
the color of the main sequence turnoff of each population.
Populations with the same turnoff colors may be of the same age and metallicity
and may be associated.  S167$-$54$-$21.5 (Sagittarius), magenta; and S341$+$57$-$22.5 (Sagittarius -- not
shown) both have turnoffs well to the blue of $g^*-r^*=0.25$, and are probably
of the same origin.  S297$+$63$-$20.0 stars (blue) and
S52$-$32$+$20.4 stars (green) also have similar turnoff colors $g^*-r^* \sim 0.26$.
The turnoff stars of the spheroid (black), S183$+$22$-$19.4 (red), S218$+$22$-$19.5 (red) and S223$+$20$-$19.4
 (red) all have similar distributions, while that of the S200$-$24$-$19.8
structure (cyan) has a bluer turnoff.
This bluer color of S200$-$24$-$19.8 might be the result of calibration errors or
too much reddening correction applied here (see Figure~\ref{tdturnoffs}).
\label{turnoffs}}

\figcaption {Same as Fig~\ref{turnoffs}, except stars with  
thick and thin disk magnitude cuts ($g^* \sim 17.5$) were selected. Note the peaks
of the histograms are in concordance except for that of S200$-$24$-$19.8 (cyan)
which is too blue in $g^*-r^*$.  This suggests that too much reddening correction
may have been applied to the data points in this region where 
estimated $E(B - V) \sim 0.12$.
\label{tdturnoffs}}

\figcaption { Sketch of the ends of stripes 37, 12, 10, and 82 in Galactic (l,b). 
Shading shows the distribution of stars in a color-magnitude selection chosen to
favor objects in the S223$+$20$-$19.4 and S200$-$24$-$19.8) structures (see \S\ref{properties}).  Black shading
denotes 600 stars per square degree or more with this selection.  Grey shading
denotes 400 stars per square degree or more.
The excess of stars near the ends of these stripes are not associated 
with known thin disk, thick disk or halo populations and may be part
of the same large structure near the Galactic anti-center at an implied
distance of 18 kpc from the Galactic center.
\label{dwarfcenter}}

\clearpage

\begin{table}
\begin{tabular}{rcclrrrc}
\multicolumn{7}{c}{Table~1 - Observing Log Summary -- Equatorial data} \\
\hline
\hline
\multicolumn{1}{c} {Run} &
\multicolumn{1}{c} {Stripe} &
\multicolumn{1}{c} {Strip} &
\multicolumn{1}{c} {Date} &
\multicolumn{1}{l} {Start RA} &
\multicolumn{1}{l} {End RA} &   
\multicolumn{1}{l} {Seeing} &
\multicolumn{1}{l} {Mult} \\
\multicolumn{1}{l} {} &
\multicolumn{1}{l} {} &
\multicolumn{1}{l} {} &
\multicolumn{1}{l} {} &
\multicolumn{1}{c} {(deg)} &
\multicolumn{1}{c} {(deg)} &   
\multicolumn{1}{l} {(arcsec)} &
\multicolumn{1}{l} {} \\
\hline
94 & 82 &N & 1998 Sep 19 & 350 & 56 & 1.7 & 1\\
125 & 82 & S & 1998 Sep 25 & 350 & 77 & 1.9 & 1\\
752 & 10 &S & 1999 Mar 21 & 145 & 233 & 1.4 & 1\\
752 & 10 &S & 1999 Mar 21 & 234 & 250 & 1.4 & 2\\
756 & 10 &N & 1999 Mar 22 & 117 & 121 & 1.4 & 2\\
756 & 10 &N & 1999 Mar 22 & 122 & 235 & 1.4 & 1\\
1350 & 37 &S & 2000 Apr  6 & 112 & 125 & 1.5 & 1\\
1402 & 37 &S & 2000 Apr 27 & 112 & 125 & 1.5 & 1\\
1450 & 37 &S & 2000 May  3 & 112 & 125 & 1.5 & 1\\
1462 & 11 & S& 2000 May  5 & 120 & 137 & 1.5 & 1\\
1752  & 82 &N & 2000 Oct  1 & 56 & 77 & 1.5 & 1\\
1755  & 82 &S & 2000 Oct  2 & 319 & 350 & 1.4 & 2\\
1907  & 11 &N & 2000 Nov 30 & 120 & 137 & 1.4 & 1\\
2125 & 12 &S & 2001 Feb 20 & 122 & 135 & 1.5 & 1\\
2126 & 12 &N & 2001 Feb 20 & 122 & 135 & 1.5 & 1\\
\hline
\end{tabular}
\end{table}

\clearpage
\begin{table}
\begin{tabular}{lrrcc}
\multicolumn{5}{c}{Table~2  - Summary of observed structures} \\
\hline
\hline
\multicolumn{1}{c} {Name} &
\multicolumn{1}{c} {Stripe} &
\multicolumn{1}{c} {Mu} &
\multicolumn{1}{c} {Turnoff} &
\multicolumn{1}{l} {Thick Disk Turnoff} \\
\multicolumn{1}{c} {S$l\pm b-g'$} &
\multicolumn{1}{c} {} &
\multicolumn{1}{c} {$^\circ$} &
\multicolumn{1}{c} {$g^*-r^*$} &
\multicolumn{1}{c} {$g^*-r^*$} \\
\hline
S6$+$41$-$20.0 & 10 & 235 & 0.30 & 0.40\\
S167$-$54$-$21.5 & 82 &  37 & 0.22 & 0.40 \\
S297$+$63$-$20.0& 10 & 190 & 0.26 & 0.40\\
8200$-$24$-$19.8 & 82 & 75 & 0.25 & 0.32\\
S223$+$20$-$19.4 & 10 & 125 & 0.28 &0.38\\
S183$+$22$-$19.4 & 37 & 135 & 0.28 &0.39 \\
S52$-$32$-$20.4 & 82 & 320 & 0.26 &0.40\\
S218$+$22$-$19.5 & 12 & 125 & 0.28 & 0.39\\
S341$+$57$-$22.5 & 10 & 213 & --- & ---\\
\hline

\end{tabular}
\end{table}

\clearpage

\setcounter{page}{1}

\plotone{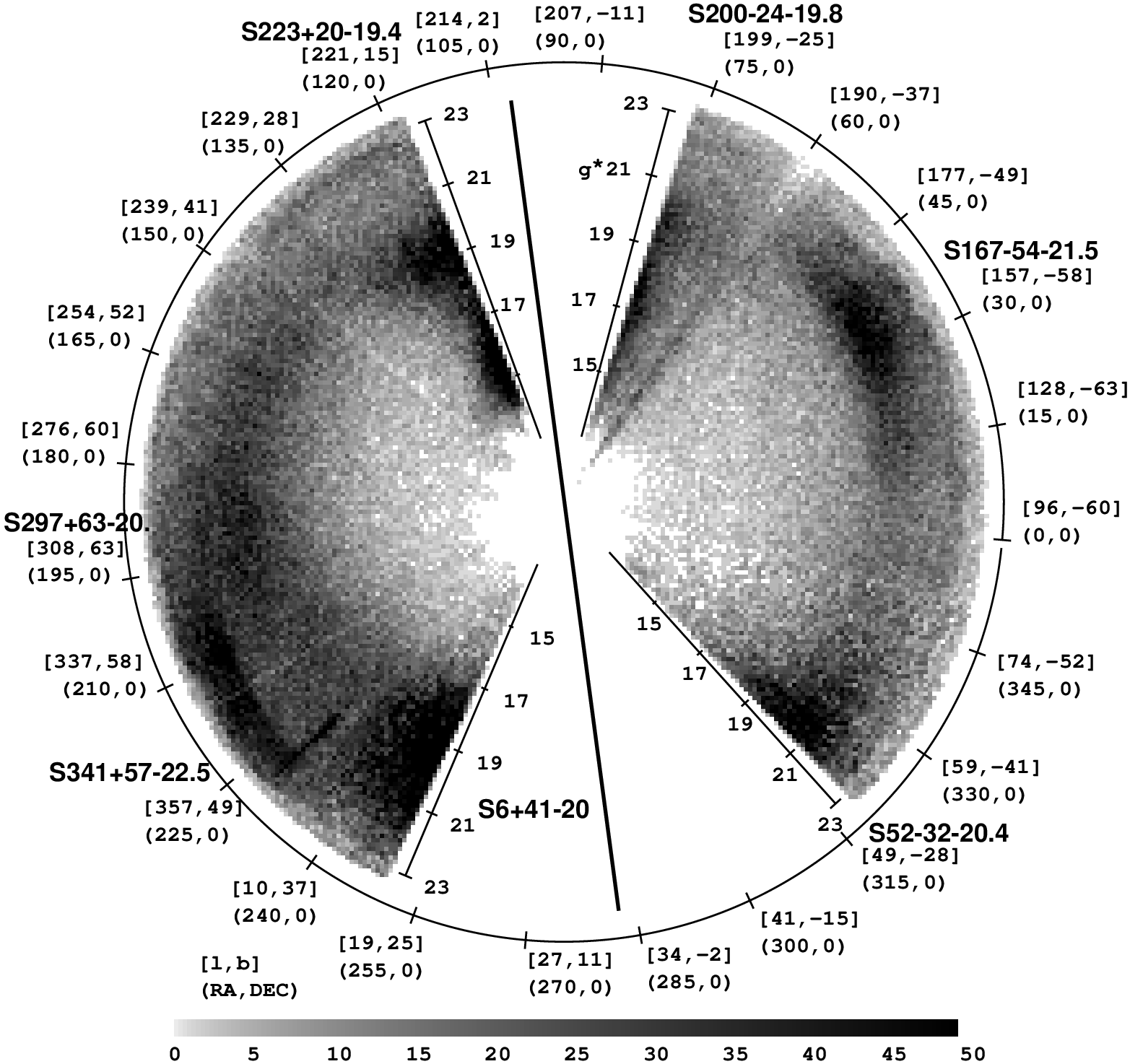}

\plotone{f2.eps}

\plotone{f3.eps}

\plotone{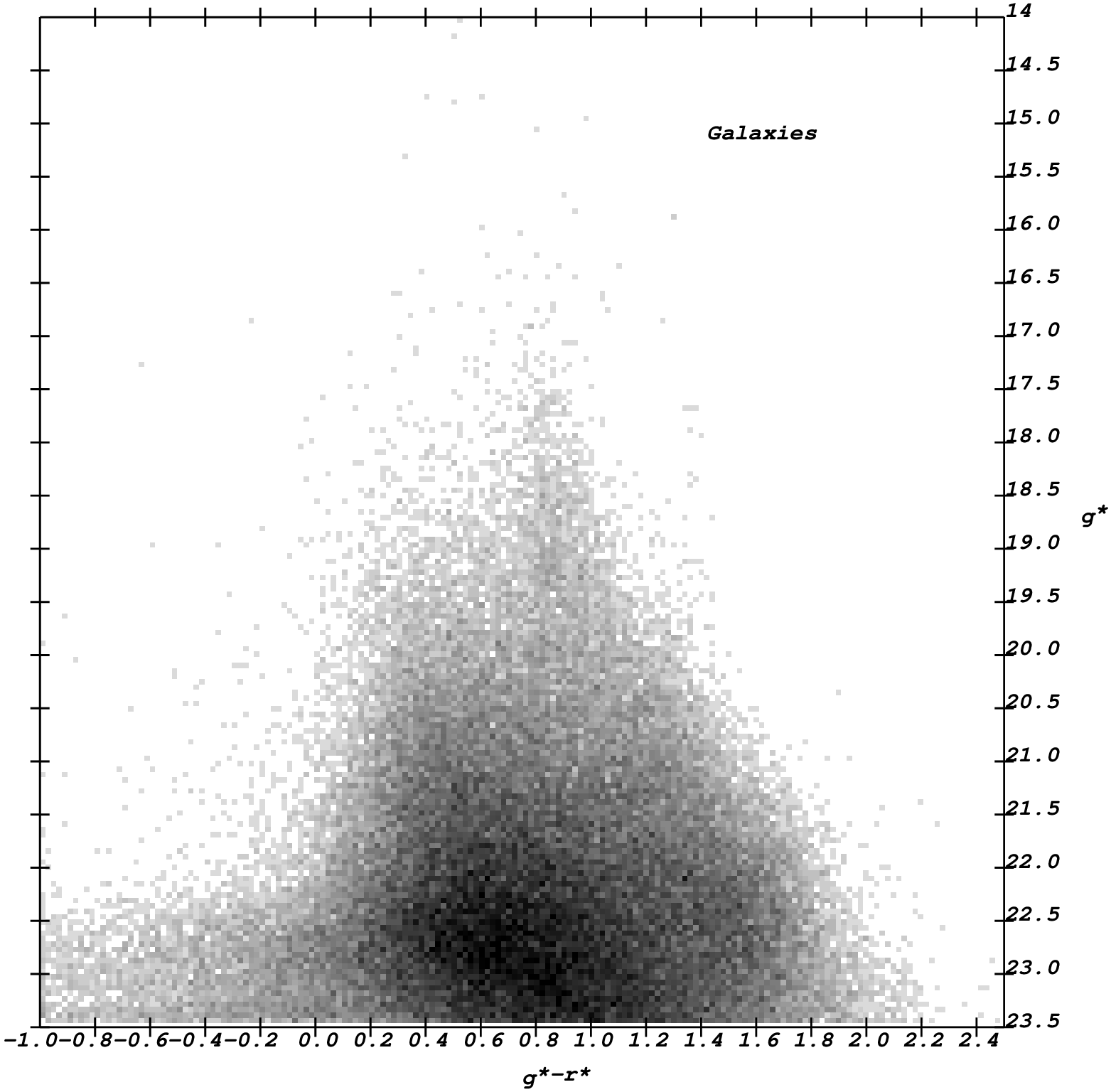}

\plotone{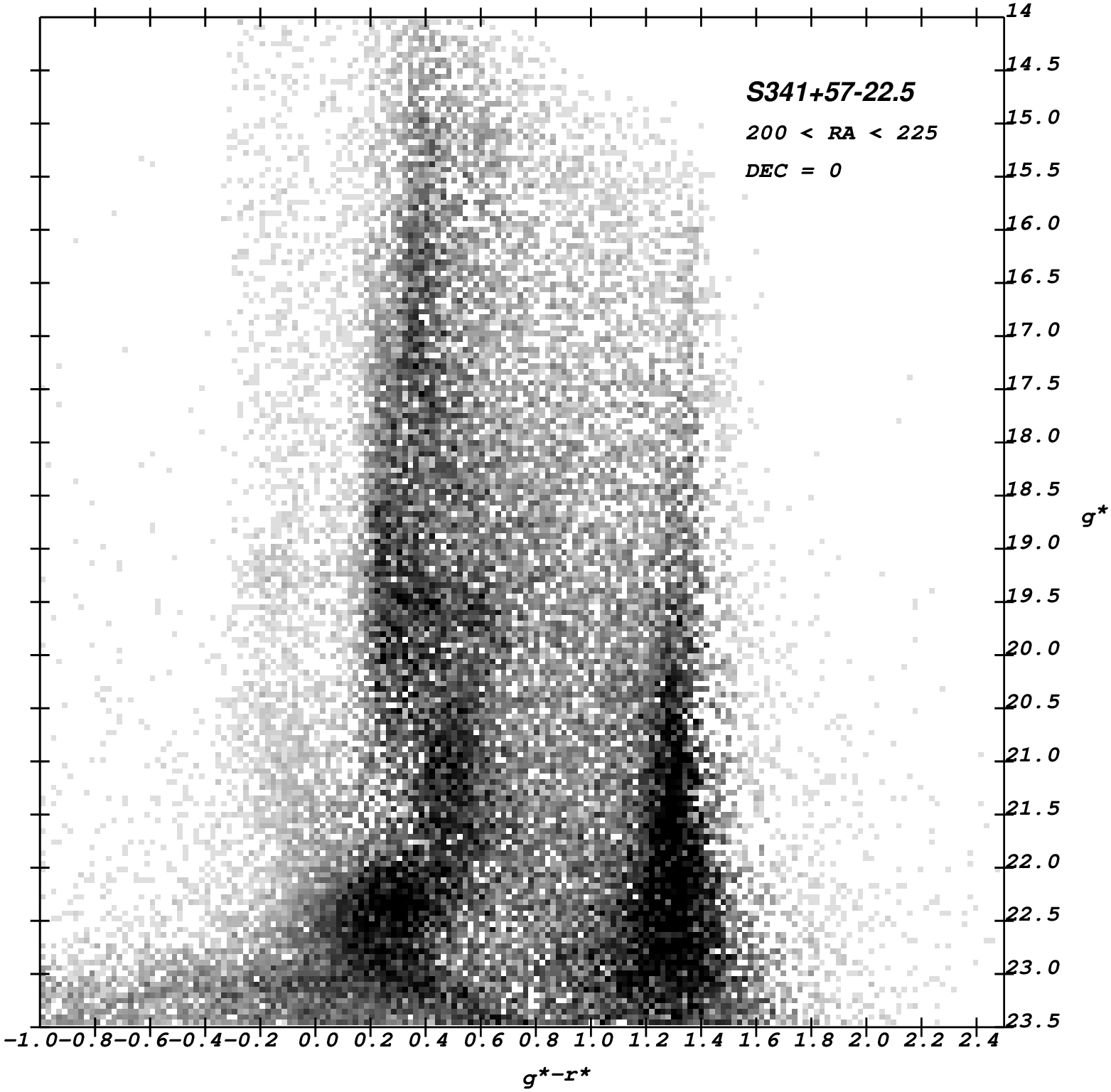}

\plotone{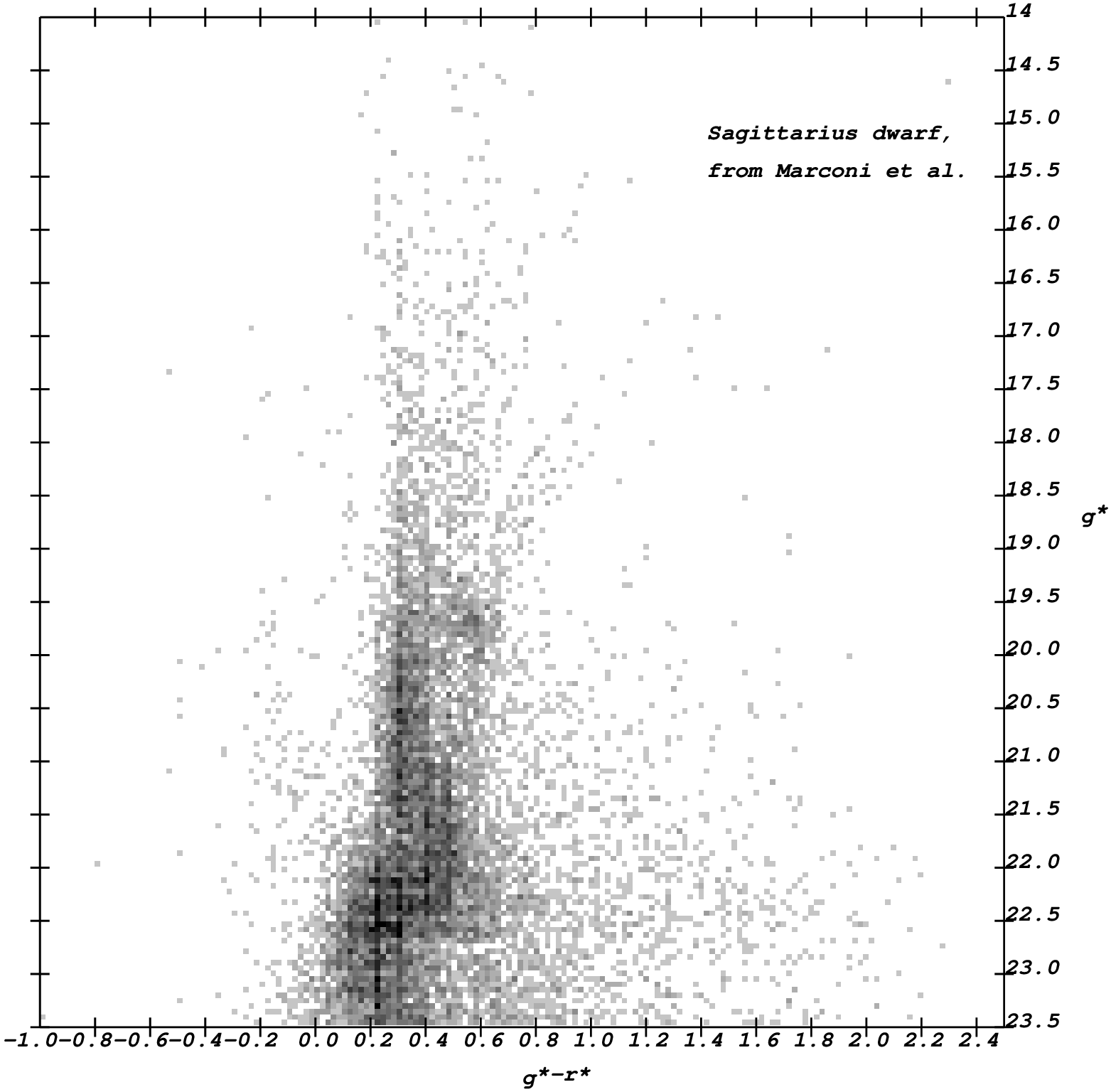}

\plotone{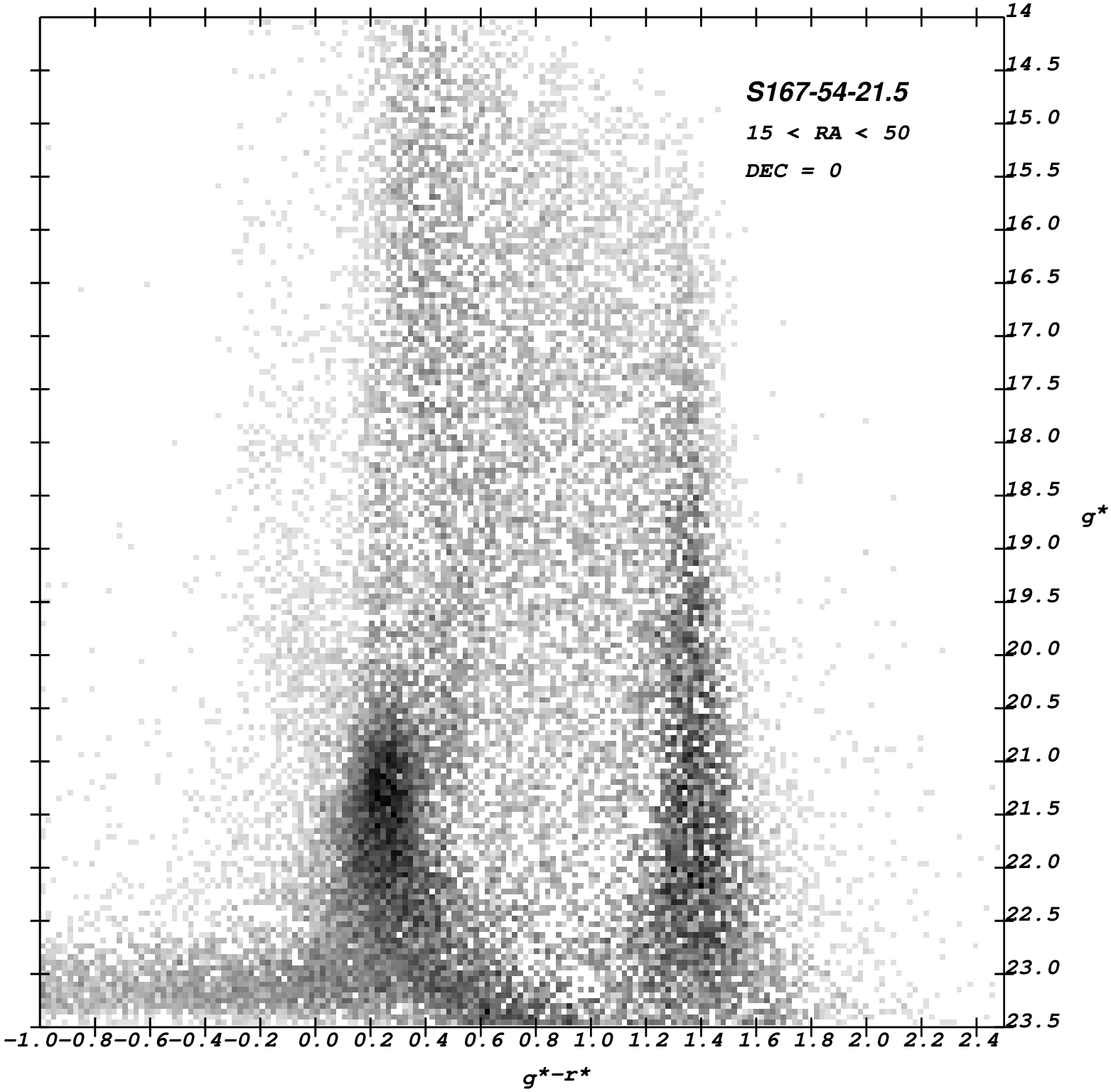}

\plotone{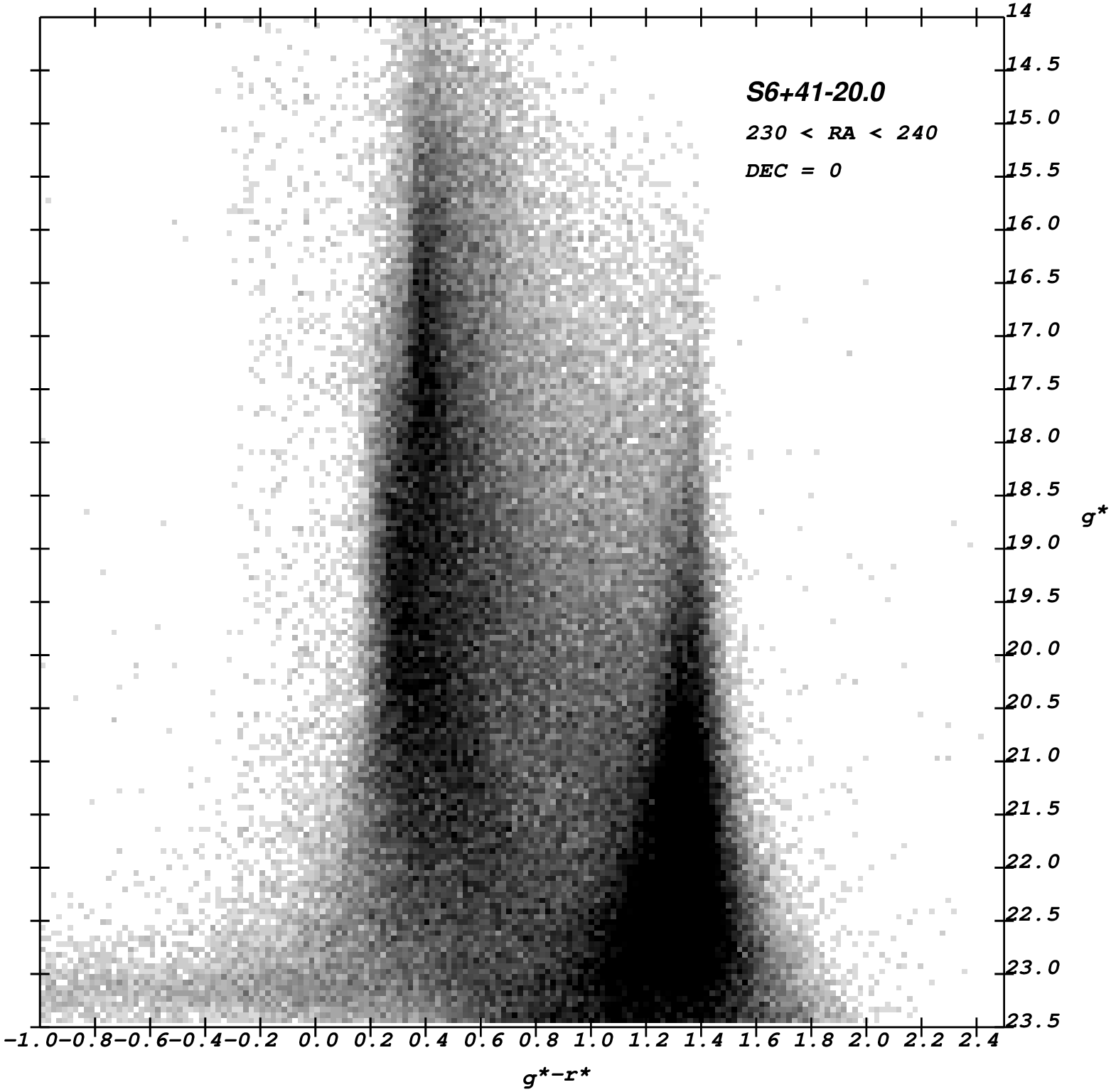}

\plotone{f9.eps}

\plotone{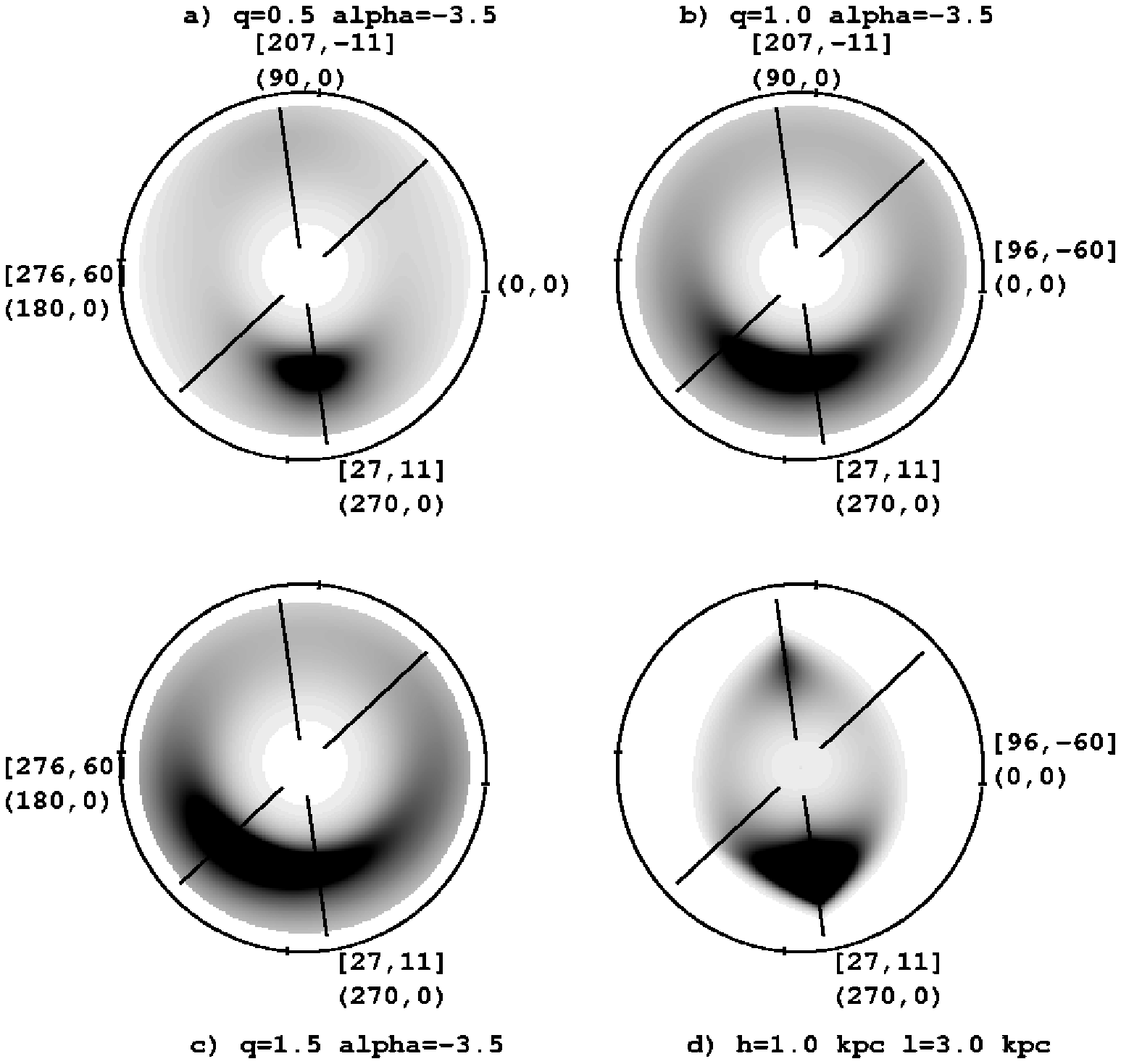}

\plotone{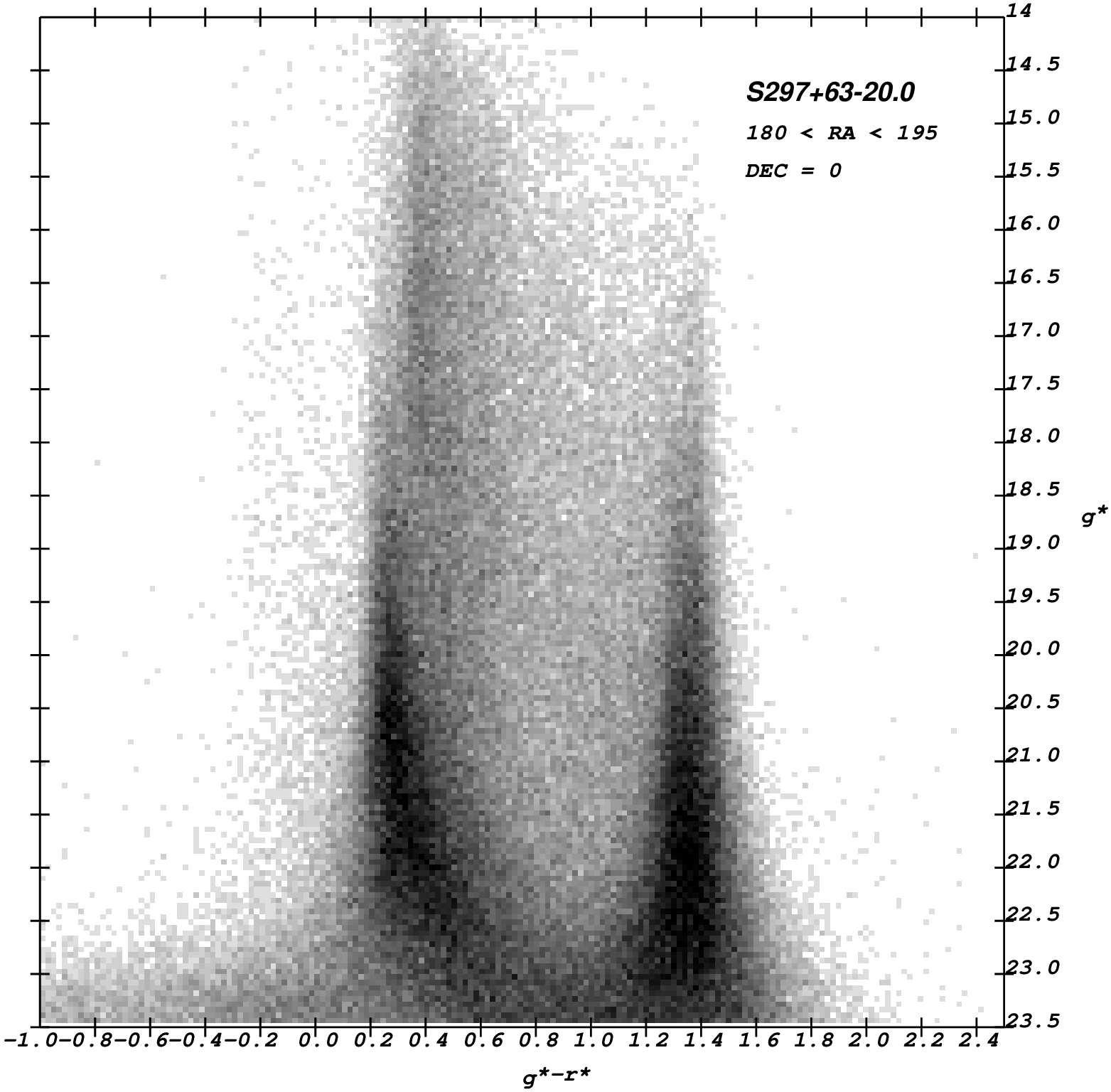}

\plotone{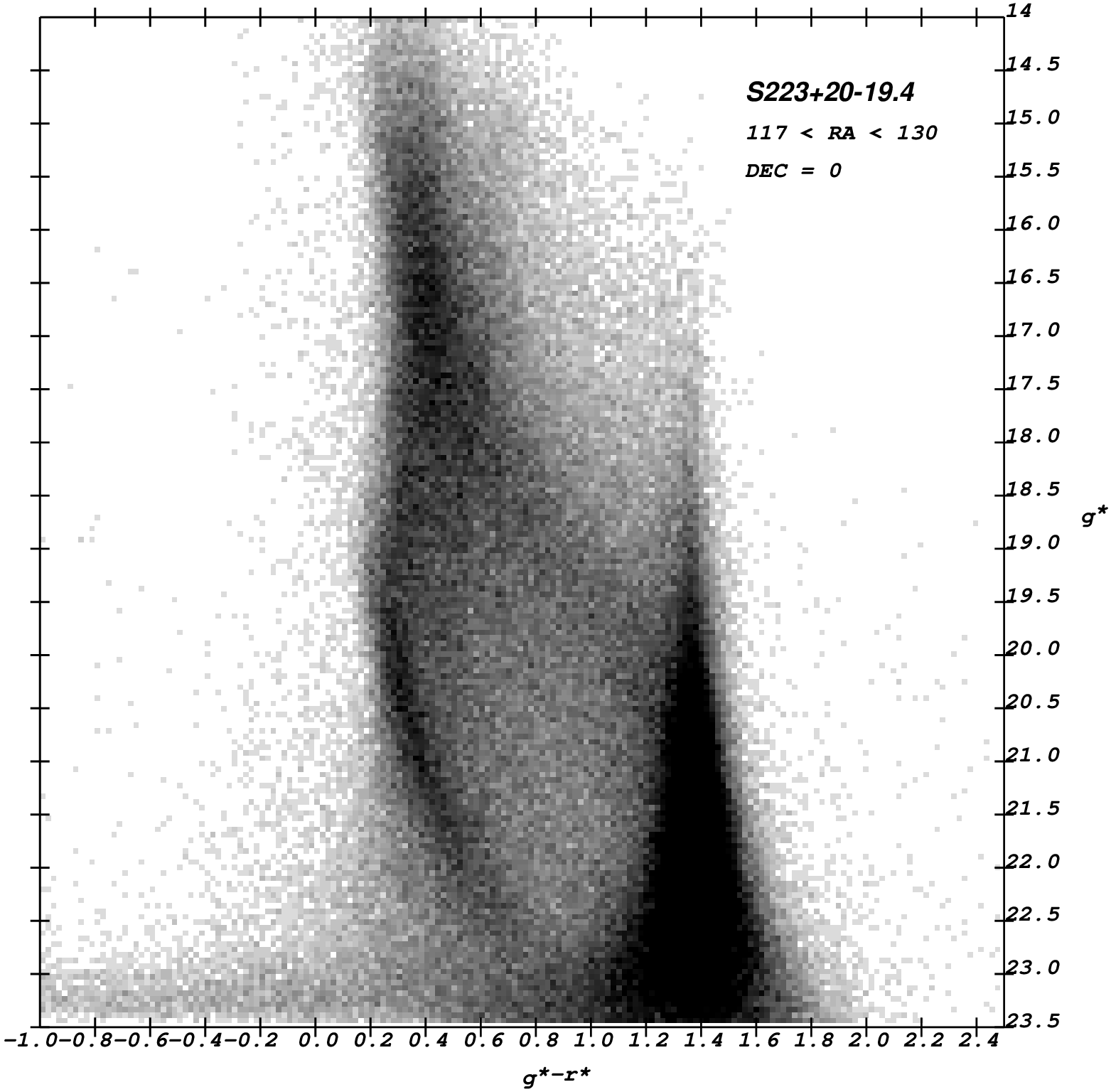}

\plotone{f13.eps}

\plotone{f14.eps}

\plotone{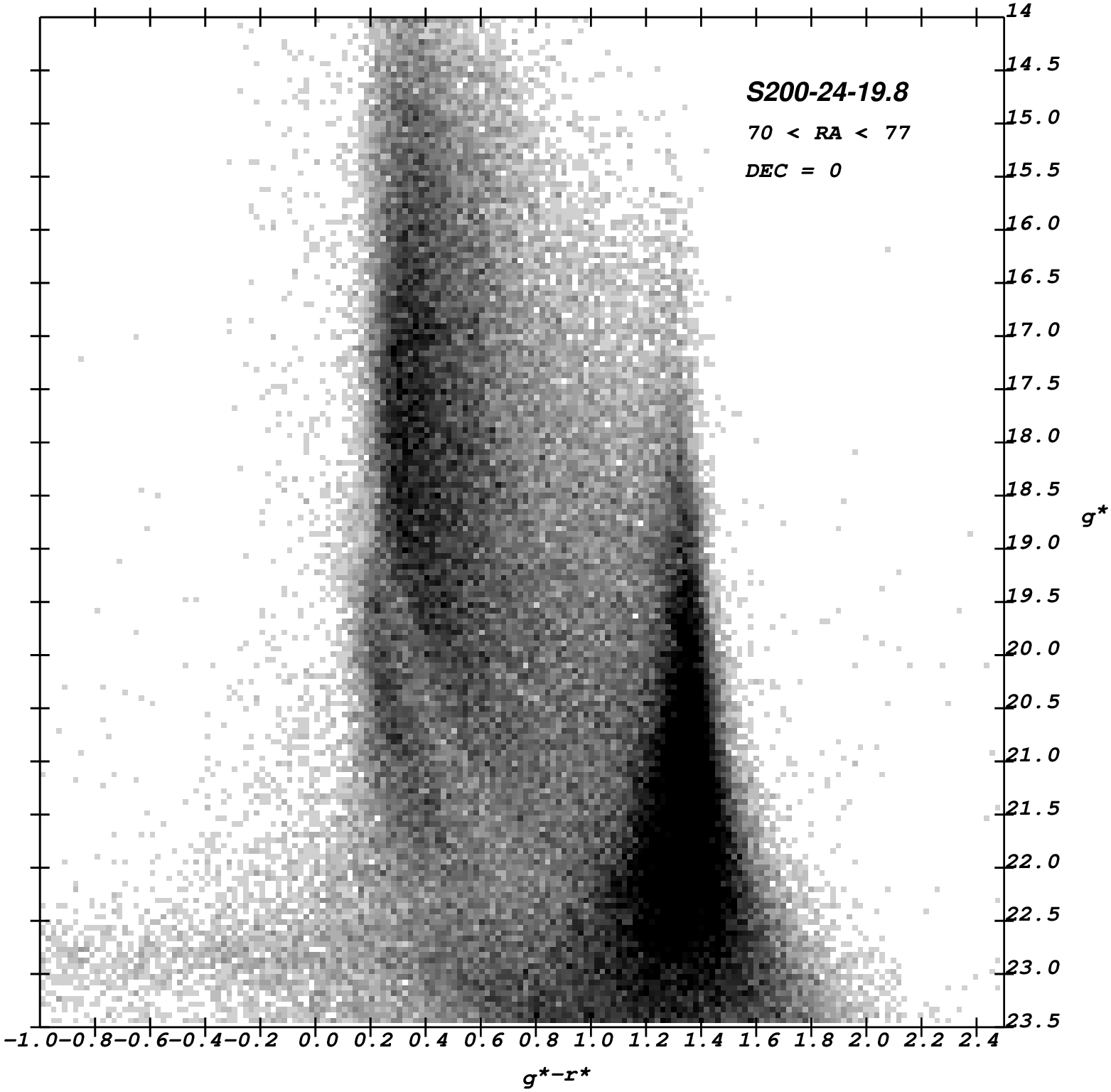}

\plotone{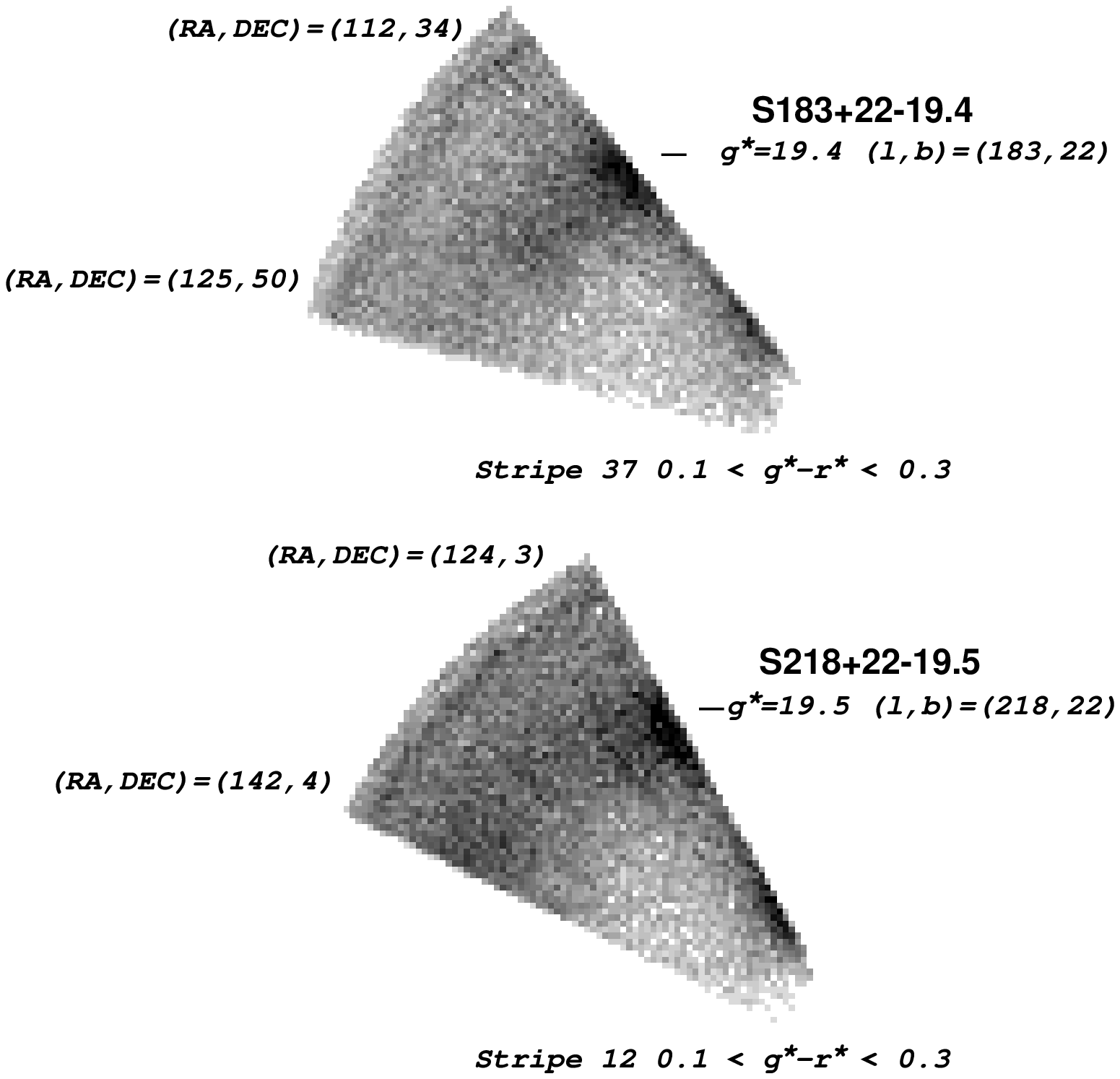}

\plotone{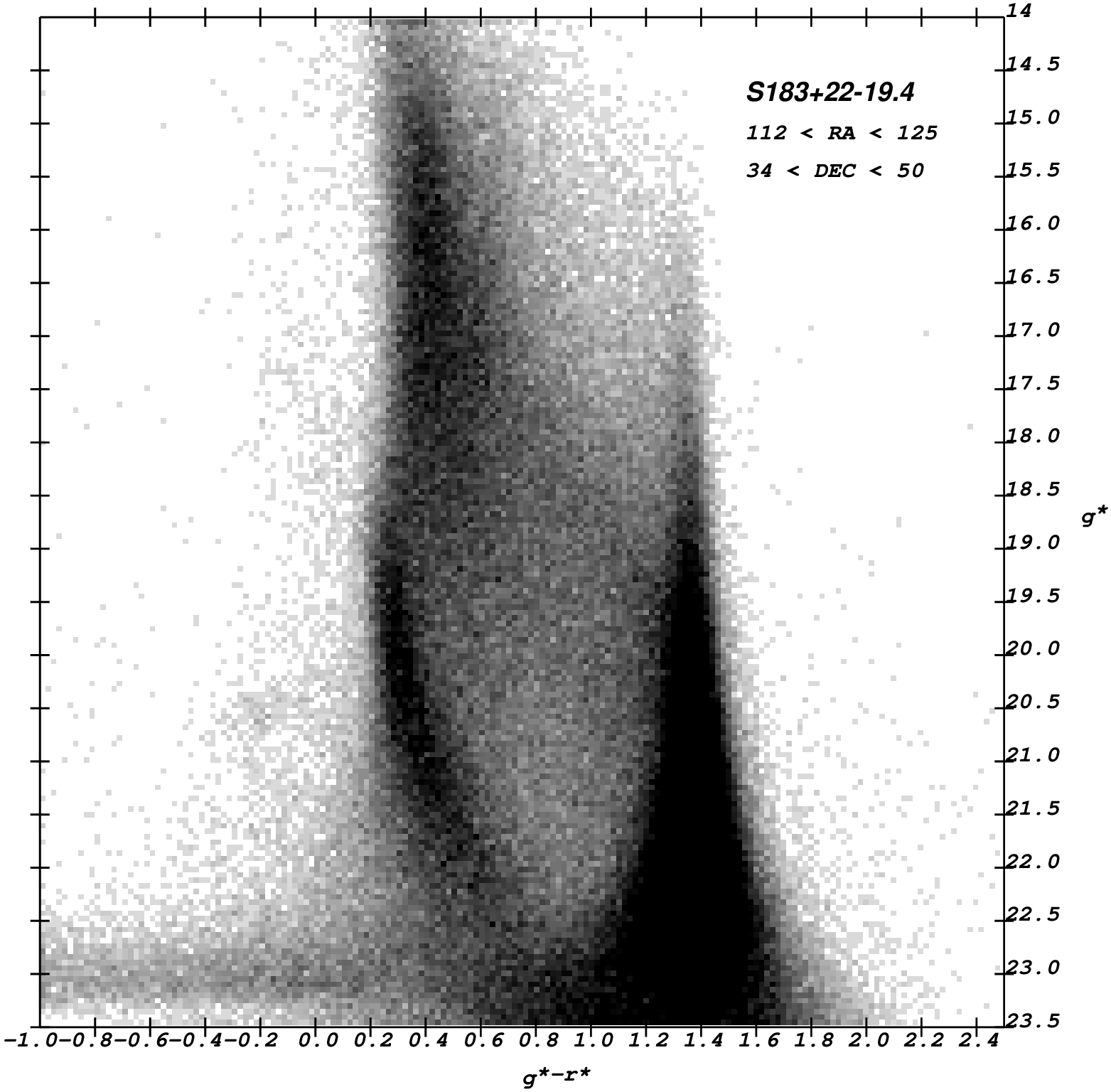}

\plotone{f18.eps}

\plotone{f19.eps}

\plotone{f20.eps}

\plotone{f21.eps}

\plotone{f22.eps}

\plotone{f23.eps}

\plotone{f24.eps}

\plotone{f25.eps}

\plotone{f26.eps}


\clearpage
\begin{thebibliography}{}
\bibitem[Alard(2001)]{a01} Alard, C. 2001, preprint astro-ph/0007013
\bibitem[Bahcall \& Soneira(1984)] {bs84} Bahcall, J. N. \& Soneira, R. M. 1984, \apjs, 55, 67
\bibitem[Binney, Gerhard, \& Spergel(1997)]{bgs97} Binney, J., Gerhard, O., \& Spergel, D.\ 1997, \mnras, 288, 365 
\bibitem[Binney \& Tremaine(1987)]{bt87} Binney, J., \& Tremaine, S. 1987, ``Galactic Dynamics,'' Princeton University Press
\bibitem[Bullock, Kravtsov, and Weinberg(2001)]{bkw01} Bullock, J. S., Kravtsov, A. V., and Weinberg, D. H. 2001, \apj, 548, 33
\bibitem[Burton \& te Lintel Hekkert(1986)]{bt86} Burton, W. B., \& te Lintel Hekkert, P. 1986, A\&AS, 65, 427
\bibitem[Buser, Rong, Karaali(1999)]{brk99} Buser, R., Rong, J., Karaali, S. 1999, A\&A, 348, 98
\bibitem[Cardelli, Clayton, \& Mathis(1989)]{ccm89} Cardelli, J. A., Clayton, G. C., \& Mathis, J. S. 1989, ApJ, 345, 245
\bibitem[Carney \& Seitzer(1993)]{cs93} Carney, B. W. \& Seitzer, P. 1993, AJ, 105, 2127
\bibitem[Chen et al.(2001)]{cetal01} Chen, B. et al. 2001, \apj, 553, 184
\bibitem[Chiba \& Beers(2000)]{cb00} Chiba, M. \& Beers, T. C. 2000, AJ, 119, 2843
\bibitem[Chiba \& Beers(2001)]{cb01} Chiba, M. \& Beers, T. C. 2001, ApJ, 549, 325
\bibitem[Cohen(1995)]{c95} Cohen, M.\ 1995, \apj, 444, 874 
\bibitem[Dohm-Palmer et al.(2001)]{detal01}Dohm-Palmer, R. C., Helmi, A., Morrison, H., Mateo, M., Olszewski, E. W., Harding, P., Freeman, K.C., Norris, J., \& Shectman, S. A. 2001, \apjl, in press (astro-ph/0105536)
\bibitem[Eggen, Lynden-Bell, and Sandage(1962)]{els62} Eggen, O. J., Lynden-Bell, D., and Sandage, A. R. 1962, \apj, 136, 748
\bibitem[Frogel(1999)]{f99} Frogel, J.~A.\ 1999, \apss, 265, 303 
\bibitem[Frogel(1988)]{f88} Frogel, J.~A.\ 1988, \araa, 26, 51 
\bibitem[Fukugita et al.(1996)]{figdss96} Fukugita, M., Ichikawa,T., Gunn, J. E., Doi, M., Shimasaku, K., Schneider, D. P. 1996, \aj, 111, 1758
\bibitem[Gilmore, Wyse, \& Kuijken(1989)]{gwk89} Gilmore, G., Wyse, R.~F.~G., \& Kuijken, K.\ 1989, \araa, 27, 555 
\bibitem[Gilmore \& Reid(1983)]{gr83} Gilmore, G., \& Reid, N. 1983, \mnras, 202, 1025
\bibitem[Guhathakurta, Choi, \& Reitzel(2000)]{gcr00} Guhathakurta, P., Choi, P.I., \& Reitzel, D. B. 2000, BAAS, 197, 3702
\bibitem[Gunn et al.(1998)]{getal98} Gunn, J. E. et al. 1998, \aj, 116, 3040
\bibitem[Hammersley, et al.(1995)]{hgmc95} Hammersley, P.~L., Garzon, F., Mahoney, T., \& Calbet, X.\ 1995, \mnras, 273, 206 
\bibitem[Helmi et al.(1999)]{hwzz99} Helmi, A., White, S. D. M., de Zeeuw, P. T., and Zhao, H. 1999, \nat, 402, 53
\bibitem[Helmi \& White(2001)]{hw01} Helmi, A., \& White, S. D. M. 2001, \mnras, 323, 529 
\bibitem[Hess(1924)]{h24} Hess, R. 1924 ``The Distribution Function of Absolute Brightness'', in  Seeliger Festschrift, ed. H. Kienle (Berlin, Springer), 265
\bibitem[Humphreys \& Larsen(1995)]{hl95} Humphreys, R.~M.~\& Larsen, J.~A.\ 1995, \aj, 110, 2183 
\bibitem[Ibata et al.(2001a)]{iils01} Ibata, R., Irwin, M., Lewis, G. F., Stolte, A. 2001a, \apj, 547, L133
\bibitem[Ibata et al.(2001b)]{ilitq01} Ibata, R., Lewis, G. F., Irwin, M., Totten, E., and Quinn, T. 2001b, \apj, 551, 294
\bibitem[Ibata and Lewis(1998)]{il98} Ibata, R. A. and Lewis, G. F. 1998, \apj, 500, 575
\bibitem[Ibata et al.(1997)]{iwgis97} Ibata, R. A., Wyse, R. F. G., Gilmore, G., Irwin, M. J., Suntzeff, N. B. 1997, \aj, 113, 634
\bibitem[Ibata, Gilmore, and Irwin(1995)]{igi95} Ibata, R. A., Gilmore, G., and Irwin, M. J. 1995, \mnras, 277, 781
\bibitem[Ibata, Gilmore, and Irwin(1994)]{igi94} Ibata, R. A., Gilmore, G., and Irwin, M. J. 1994, \nat, 370, 194
\bibitem[Ivezi\'{c} et al.(2000)]{ietal00} Ivezi\'{c}, Z., et al. 2000, \aj, 120, 963
\bibitem[Jiang and Binney(2000)]{jb00} Jiang, I. and Binney, J. 2000, \mnras, 314, 468
\bibitem[Johnston et al.(1999a)]{jmsrk99} Johnston, K. V., Majewski, S. R., Siegel, M. H., Reid, I. N., Kunkel, W. E. 1999a, \aj, 118, 1719
\bibitem[Johnston et al.(1999b)]{jzsh99} Johnson, K. V., Zhao, H., Spergel, D. N., and Hernquist, L. 1999b, \apjl, 512, L109
\bibitem[Johnston, Sigurdsson, and Hernquist(1999)]{jsh99} Johnston, K. V., Sigurdsson, S., Hernquist, L. 1999, \mnras, 302, 771
\bibitem[Johnston, Hernquist, and Bolte(1996)]{jhb96} Johnston, K. V., Hernquist, L., and Bolte, M. 1996, \apj, 465, 278
\bibitem[Johnston, Spergel and Hernquist(1995)]{jsh95} Johnston, K. V., Spergel, D. N., Hernquist, L. 1995, \apj, 451, 598
\bibitem[Kinman, Suntzeff, \& Kraft(1994)]{ksk94} Kinman, T. D., Suntzeff, N. B., \& Kraft, R. P. 1994, \aj, 108, 1722
\bibitem[Klypin et al.(1999)]{klypin99} Klypin, A. A., Kravtsov, A. V., Valenzuela, O., \& Prada, F. 1999, \apj, 522, 82
\bibitem[Kerber, Javiel, Santiago(2001)]{kjs01} Kerber, L. O., Javiel, S. C., \& Santiago, B. X. 2001, A\&A, 365, 424
\bibitem[Larsen \& Humphreys(1996)]{lh96} Larsen, J. A., \& Humphreys, R. M. 1996, \aj, 468, 99
\bibitem[Layden \& Sarajedini(2000)]{ls00} Layden, A., \& Sarajedini, A. 2000, \aj, 119, 1760
\bibitem[Lupton, Gunn, \& Szalay(1999)]{lgs99} Lupton, R. H., Gunn, J. E., \& Szalay, A. S. 1999, AJ, 118, 1406 
\bibitem[Lupton et al.(2001)]{l01} Lupton, R. H., et al., \aj, in preparation
\bibitem[Lynden-Bell and Lynden-Bell(1995)]{ll95} Lynden-Bell, D., and Lynden-Bell, R. M. 1995, \mnras, 275, 429
\bibitem[Majewski et al.(1999)]{mskrjtlp99} Majewski, S. R., Siegel, M. H., Kunkel, W. E., Reid, I. N., Johnston K. V., Thompson, I. B., Landolt, A. U., and Palma, C. 1999, \aj, 118, 1709
\bibitem[Majewski, Munn, and Hawley(1996)]{mmh96} Majewski, S. R., Munn, J., A., and Hawley, S. L. 1996, \apjl, 459, L73
\bibitem[Majewski(1994)]{m94} Majewski 1994, \apj, 431, 1
\bibitem[Majewski(1993)]{m93} Majewski, S.~R.\ 1993, \araa, 31, 575 
\bibitem[Marconi et al.(1998)]{mbcimpp98} Marconi, G., Buonanno, R., Castellani, M., Iannicola, G., Molaro, P., Pasquini, L., and Pulone, L. 1998, A \& A, 330, 453
\bibitem[Martinez-Delgado et al.(2001)]{magc01} Martinez-Delgado, D., Aparicio, A., Gomez-Flechoso, M. Angeles, \& Carrera, Ricardo 2001, \apj, 549, L199
\bibitem[Mendez \& van Altena(1998)]{mv98} Mendez, R.~A.~\& van Altena, W.~F.\ 1998, \aap, 330, 910 
\bibitem[Morrison, Flynn, \& Freeman(1990)]{mff90} Morrison, H. L., Flynn, C., \& Freeman, K. C. 1990, \aj, 100, 1191
\bibitem[Moore et al.(1999)]{moore99} Moore, B., Ghigna, S., Governato, F., Lake, G., Quinn, T., Stadel, J., \& Tozzi, P. 1999, \apj, 524, L19
\bibitem[Norris(1999)]{n99} Norris, J.~E.\ 1999, \apss, 265, 213 
\bibitem[Norris(1994)]{n94} Norris, J. E. 1994, \apj, 431, 645
\bibitem[Odenkirchen et al.(2001)]{oetal01} Odenkirchen, M. et al. 2001, \apjl, 548, L165
\bibitem[Ojha et al.(1996)]{obrcm96} Ojha, D. K., Bienayme, O., Robin, A. C., Creze, M., Mohan, V. 1996, A\&A 311, 456
\bibitem[Pier et al.(2001)]{pmk01} Pier, J., Munn, J. A., Kent, S. M., et al. 2001, \aj, in preparation
\bibitem[Reid \& Majewski(1993)]{rm93} Reid, N. \& Majewski, S. R. 1993, ApJ, 409, 635
\bibitem[Reyle \& Robin(2001)]{rr01} Reyle, C., \& Robin, A. C. 2001, A\&A, 373, 886
\bibitem[Robin et al.(1996)]{rhcob96} Robin, A. C., Haywood, M., Creze, M., Ojha, D. K. \& Bienayme, O. 1996, A\&A, 305, 125
\bibitem[Schlegel, Finkbeiner, \& Davis(1998)]{sfd98} Schlegel, D.J., Finkbeiner, D.P., \& Davis, M. 1998, ApJ, 500, 525
\bibitem[Scranton, Johnston, \& Lupton(2001)]{sjl01} Scranton, R., Johnston, D., \& Lupton, R. H. 2001, AJ, submitted.
\bibitem[Smith et al., in preparation]{setal} Smith et al., in preparation
\bibitem[Stoughton et al.(2001)]{setal01} Stoughton, C., et al. 2001, \aj, in press
\bibitem[Vivas et al.(2001)]{v01} Vivas, A.~K.~et al.\ 2001, \apjl, 554, L33 
\bibitem[Wyse(1999)]{w99} Wyse, R.~F.~G.\ 1999, Baltic Astronomy, 8, 593 
\bibitem[Yanny et al.(2000)]{ynetal00} Yanny, B., Newberg, H. J., et al. 2000, \apj, 540, 825 (Paper I)
\bibitem[York et al.(2000)]{yetal00} York, D.G.  et al. 2000, AJ, 120, 1579
\end{thebibliography}
\end{document}